\documentclass{aero}

\makeatletter
\def\@ADBIOphotodepth{1.35in}
\def\@ADBIOhangdepth{1.35in}
\makeatother

\usepackage{tabularx}
\usepackage[short]{optidef}
\usepackage{graphicx}
\usepackage{float}
\usepackage[section]{placeins}

\ADsetup{
title    = {B-spline shaping for low-thrust interplanetary rendezvous},
author   = {Julio C. Sanchez$^{1}$\cor{}},
address  = {
  1. Department of Aerospace Engineering and Fluid Mechanics, Universidad de Sevilla, Sevilla, 41092, Spain\sep
},
email    = {jsanchezm@us.es},
abstract = {
	Rapid preliminary design of low-thrust interplanetary rendezvous
	trajectories requires parameterizations that combine computational
	efficiency, flexibility, and sufficient solution quality. This work
	presents a shape-based method in which the heliocentric cylindrical
	coordinates are parameterized using clamped B-splines. The differential
	flatness of the low-thrust cylindrical dynamics is exploited to recover the
	control acceleration algebraically from the shaped trajectory and its
	derivatives. The clamped B-spline parameterization satisfies the endpoint
	position and velocity conditions analytically, reducing the original
	fixed-time, minimum-$\Delta V$ problem to a
	finite-dimensional nonlinear program in the free interior B-spline control
	points. The formulation therefore requires no numerical propagation of
		the equations of motion; numerical quadrature is used only to evaluate
		the objective function. Numerical campaigns are conducted for
low-thrust rendezvous transfers from Earth to Mars, Mercury, the near-Earth
asteroid 1989~ML, and comet Tempel~1, and the results are compared
		with high-order hodographic shaping. Across the tested fine-grid campaigns
		and among the finite solutions retained, all B-spline configurations
		reduce the mean, median, and minimum accumulated velocity increment
		relative to the hodographic benchmark. In particular, a
	ten control-point quintic B-spline is identified as a favorable compromise between
	solution quality and computational efficiency, while higher-dimensional
	parameterizations are effective for refining selected transfer
	opportunities.
  },
keywords = {
    Low-thrust trajectory optimization \sep
  B-spline shaping \sep
  Shape-based methods \sep
  Interplanetary rendezvous
},
}

\ADsetup{
	name = {},
	url  = {},
}

\begin{document}

\maketitle  

\section{Introduction}
\label{sec:introduction}

The demand for fuel-efficient, high-capability space missions has motivated the widespread adoption of low-thrust propulsion systems, particularly electric thrusters, as alternatives to conventional chemical propulsion. Their substantially higher specific impulse reduces the propellant mass required to achieve a prescribed velocity increment, thereby enabling larger payloads or more demanding mission architectures. The viability of electric propulsion has been demonstrated by several missions. Deep Space~1 validated solar electric propulsion as a primary propulsion system in deep space \cite{Rayman2000}; SMART-1 used a Hall-effect thruster for lunar orbit raising \cite{Racca2002}; Dawn employed ion propulsion to orbit both Vesta and Ceres \cite{Brophy2011}; and Hayabusa missions used electric propulsion to support sample-return from the near-Earth asteroids Itokawa and Ryugu, respectively \cite{Kuninaka2007,Tsuda2013}. Unlike chemical maneuvers, which can often be approximated as instantaneous velocity changes, low-thrust transfers require continuous or intermittent thrusting over extended periods and may involve several orbital revolutions. Their mission design therefore constitutes a challenging trajectory-optimization problem.

Continuous-thrust trajectory design is traditionally formulated as an optimal control problem. Indirect methods apply Pontryagin's minimum principle \cite{Pontryagin1962}, producing a nonlinear two-point boundary-value problem in the state and costate variables \cite{Bryson1975}. For bounded-thrust minimum-fuel problems, the optimal throttle commonly exhibits a bang-bang or bang-off-bang structure governed by a switching function \cite{Russell2007,Prussing2010}. The resulting shooting problem may be highly sensitive to the initial costate estimate, particularly for long-duration and multi-revolution transfers \cite{Morante2021}. Several techniques have been developed to alleviate this difficulty. An enhanced smoothing technique combined with continuation was proposed to regularize the switching structure \cite{TaheriSmoothing2016}. Reference \cite{Ottesen2024} developed a direct-to-indirect mapping that uses impulsive solutions to initialize indirect low-thrust optimization. Indirect optimization has also been integrated with successive convex programming to improve costate initialization \cite{Tang2018}. Other applications include the indirect optimization of fuel-optimal Earth--Moon transfers \cite{PerezPalau2018}, costate mapping between different coordinate sets \cite{Taheri2021}, and power-limited asteroid rendezvous trajectories \cite{Wang2022}. Although these developments improve convergence, indirect methods generally continue to require careful initialization and problem-specific numerical strategies.

Direct methods instead transcribe the continuous optimal control problem into a finite-dimensional nonlinear program \cite{Hargraves1987,Betts1998}. They are generally more robust to initialization than indirect methods, although the resulting solution approximates the original continuous problem through a finite-dimensional discretization. Direct collocation represents the state and control histories using polynomial functions and enforces the equations of motion at selected points \cite{Topputo2014}. Pseudospectral methods follow the same general principle using global orthogonal-polynomial approximations and specialized collocation nodes \cite{Garg2010}. For low-thrust rendezvous, rapid shape-based trajectory generation has been combined with Radau pseudospectral optimization, using the shaped trajectory to initialize a higher-fidelity refinement \cite{Li2014}. Direct transcription has also been used to generate time-optimal interplanetary transfers for electric solar wind sails as part of model-predictive guidance architectures \cite{Urrios2024}. Nevertheless, direct transcription methods may produce large nonlinear programs and still require a suitable initial trajectory, which can be difficult to construct for multi-revolution transfers. Moreover, detecting feasible low-thrust transfers can be cumbersome, and impulsive screening followed by continuous-thrust refinement may be necessary \cite{Mereta2018}. Successive convexification provides another alternative by solving a sequence of convex subproblems about a reference trajectory \cite{Bernardini2024}.

Shape-based methods provide an efficient alternative for preliminary trajectory design. Rather than discretizing the complete state and control histories, they parameterize selected trajectory or velocity variables using a comparatively small set of basis functions. Early work introduced the exponential-sinusoid representation for low-thrust gravity-assist trajectory design \cite{Petropoulos2004}. This concept was subsequently extended by optimizing free coefficients in the shaping functions \cite{NovakVasile2011}. Hodographic shaping, in which the cylindrical velocity components are parameterized using analytical basis functions, was proposed in \cite{Gondelach2015}. That formulation distinguishes between low-order solutions, whose coefficients are fully determined by the boundary conditions, and high-order solutions containing additional coefficients that can be optimized. A three-dimensional shaping approach based on finite Fourier series was developed in \cite{TaheriShape2016}. B\'ezier curves have also been used to parameterize heliocentric cylindrical coordinates for the rapid generation of three-dimensional low-thrust interplanetary trajectories \cite{Fan2021}. Reference~\cite{Wu2022} proposed a rendezvous-shaping method based on piecewise cubic splines of modified equinoctial elements. Their formulation satisfies the rendezvous boundary conditions analytically and can retain free spline parameters for optimization, including the treatment of a thrust-magnitude constraint. These methods provide rapid solutions, but their performance may depend on the selection or construction of problem-specific shaping variables and basis functions. More recently, shape-based methods have been shown to provide effective initial guesses for pseudospectral methods \cite{Guo2025}.

This work proposes a B-spline shaping method for fixed-time, minimum-$\Delta V$ low-thrust rendezvous. The heliocentric cylindrical coordinates are represented directly using clamped B-splines \cite{DeBoor1972}. Because the cylindrical low-thrust dynamics are differentially flat with respect to the cylindrical coordinates, the control acceleration can be recovered algebraically from the shaped trajectory and its first two derivatives  \cite{Fliess1995}. The endpoint position and velocity conditions are imposed through the first and last B-spline control points, leaving only the interior control points as optimization variables. Consequently, the original continuous optimal control problem is reduced to a bounded finite-dimensional nonlinear program. The equations of motion and boundary conditions are satisfied by construction and therefore need not be imposed as separate transcription constraints or evaluated by numerically propagating the dynamics. B-splines \cite{DeBoor1972} offer several properties that are attractive for shape-based trajectory design. They provide a compact and smooth representation, support analytical differentiation, and possess local support, such that modifying one control point affects only a limited portion of the trajectory. Clamped B-splines can also enforce endpoint position and velocity conditions analytically. Unlike reference~\cite{Wu2022}, who constructs piecewise cubic splines of modified equinoctial elements, the present method parameterizes the three cylindrical position coordinates directly using a standard clamped B-spline basis and exploits their explicit flatness mapping to the control acceleration. The degree and number of control points can therefore be varied systematically without deriving a new piecewise shaping construction. Related B-spline and differential-flatness parameterizations have also been applied to spacecraft attitude path planning and relative-motion control \cite{Louembet2009, Louembet2011, Sanchez2020_AA}. To the author's knowledge, this specific combination of clamped B-spline cylindrical-position shaping and the differential flatness of the heliocentric dynamics has not previously been investigated for low-thrust interplanetary rendezvous.

The main contributions of this work are twofold. First, a shape-based
formulation is developed by combining a clamped B-spline parameterization
of the cylindrical coordinates with the differential flatness of the
low-thrust dynamics. This combination allows the control acceleration to
be recovered algebraically from the shaped trajectory and its derivatives,
while the endpoint position and velocity conditions are enforced by
construction through the clamped B-spline. Consequently, the continuous
fixed-time, minimum-$\Delta V$ problem is reduced to a bounded
finite-dimensional nonlinear program in the free interior B-spline control
points,
without requiring numerical propagation of the equations of motion or the
introduction of dynamic collocation constraints; numerical quadrature is
required only for the objective function. Furthermore, the
flexibility of the trajectory representation can be adjusted systematically
through standard B-spline hyperparameters, primarily the polynomial degree
and number of control points, instead of manually constructing
target-specific combinations of analytical basis functions. Second, the proposed method is systematically benchmarked against the high-order hodographic shaping method of \cite{Gondelach2015} through
fine-grid rendezvous campaigns from Earth to Mars, Mercury, the near-Earth
asteroid 1989~ML, and comet Tempel~1. In particular, the quintic
B-spline with ten control points reduces the median $\Delta V$ by
approximately 22--45\%, depending on the target, while requiring
approximately 30--33\% less computation time per transfer attempt than the
hodographic benchmark. This configuration therefore provides an
advantageous compromise between optimality and computational
efficiency. Configurations with a larger number of control points provide
further reductions in the minimum $\Delta V$ when solution quality is
prioritized over computational cost.

The remainder of this paper is organized as follows. Section~2 formulates the low-thrust interplanetary rendezvous optimization problem. Section~3 develops the proposed B-spline shaping method. Section~4 describes the numerical implementation and hodographic benchmark. Section~5 presents the numerical results and discusses optimality and computational efficiency. Finally, Section~6 summarizes the principal conclusions.

\section{Problem formulation for low-thrust interplanetary rendezvous}

This work considers a fixed-time, minimum-$\Delta V$ low-thrust rendezvous problem for a prescribed number of revolutions. The objective is to minimize the time integral of the control-acceleration magnitude. Spacecraft mass is omitted so that the development can focus on the shape-based trajectory formulation. The state is expressed in heliocentric cylindrical coordinates $(r,\theta,z)$: \(r\) is the distance from the Sun to the spacecraft projection onto the ecliptic plane, \(\theta\) is the corresponding azimuthal angle, and \(z\) is the axial displacement from the ecliptic. Under these assumptions, the optimal control problem is
\begin{mini}
	{\mathbf{u}}
	{\Delta V = \int_{t_0}^{t_f} \lVert\mathbf{u}\rVert\, \mathrm{d}t,}
	{\label{prob:main}}{}
	\addConstraint{\dot{r} = v_r,\quad \dot{\theta} = \frac{v_\theta}{r},\quad \dot{z}= v_z}
	\addConstraint{\dot{v}_r= \frac{v_\theta^2}{r} - \frac{\mu r}{s^{3}} + u_r}
	\addConstraint{\dot{v}_\theta= -\frac{v_r v_\theta}{r} + u_\theta}
	\addConstraint{\dot{v}_z= -\frac{\mu z}{s^{3}} + u_z}
	\addConstraint{r(t_0)=r_0,\quad \theta(t_0)=\theta_0,\quad z(t_0)=z_0}
	\addConstraint{r(t_f)=r_f,\quad \theta(t_f)=\theta_f+2\pi N,\quad z(t_f)=z_f}
	\addConstraint{v_r(t_0)=v_{r_0},\quad v_{\theta}(t_0)=v_{\theta_0},\quad v_z(t_0)=v_{z_0}}
	\addConstraint{v_r(t_f)=v_{r_f},\quad v_{\theta}(t_f)=v_{\theta_f},\quad v_z(t_f)=v_{z_f}.}
\end{mini}
The variable $s=\sqrt{r^2+z^2}$ is the orbital radius with respect to the Sun. The boundary conditions $(r_0,\theta_0,z_0,v_{r_0},v_{\theta_0},v_{z_0})$ and $(r_f,\theta_f,z_f,v_{r_f},v_{\theta_f},v_{z_f})$ are obtained from the ephemerides of the departure and target bodies at times $t_0$ and $t_f$, respectively. The transfer time $T=t_f-t_0$ and the number of complete revolutions $N\in\mathbb{N}$ (including zero) are fixed beforehand to explore different multi-revolution transfers. The terminal azimuth $\theta_f$ obtained from the target ephemeris is expressed as
the first equivalent prograde angle not smaller than the initial
azimuth. For a solution containing \(N\) complete revolutions, \(N\)
additional full turns are then added to this angle. The control acceleration is left unconstrained to
retain a general shape-based formulation. Accordingly, the solutions
are not guaranteed to satisfy the acceleration capability of a
particular low-thrust propulsion system. Section~\ref{sec5} presents
a posteriori comparisons of the sampled maximum control norm,
\(u_{\mathrm{max}}\), and the resulting \(\Delta V\). 

\section{B-spline shaping for low-thrust trajectory optimization}\label{sec:3}

This section presents a B-spline shaping method for low-thrust trajectory
optimization. First, the differential flatness of the cylindrical
low-thrust dynamics is introduced and used to recover the control
acceleration algebraically from the cylindrical coordinates and their
derivatives. The heliocentric cylindrical coordinates are subsequently
parameterized using B-splines, and the relevant spline properties are
summarized. Next, the control acceleration is expressed explicitly as a
function of the B-spline parameters. Finally, the formulation is reduced to
a nonlinear program that does not require numerical propagation of the
equations of motion and in which the boundary conditions are satisfied by
construction through a clamped B-spline.

\subsection{Differential flatness of the cylindrical low-thrust dynamics}

A key feature of the proposed shape-based method is the transformation of the continuous optimal control problem~\eqref{prob:main} into a nonlinear program that requires no numerical propagation of the dynamics. To this end, the differential flatness property of the cylindrical low-thrust dynamics is exploited. A system is differentially flat if its state and control can be expressed algebraically in terms of a flat output and a finite number of its derivatives \citep{Fliess1995}. In this problem, the flat output is the vector of cylindrical coordinates, \(\mathbf{q}=[r,\theta,z]^T\). The corresponding flatness mapping is valid on the nonsingular domain \(r>0\), where the cylindrical angle is well defined. The property can be verified directly by first using the kinematics to express the cylindrical velocity $\mathbf{v}$ as a function of \(\mathbf{q}\) and \(\dot{\mathbf{q}}\):
\begin{equation}
	\mathbf{v}=[
		\dot{r},\,
		r\dot{\theta},\,
		\dot{z}]^T.
	\label{eq:flatness1}
\end{equation}
Then, by using the low-thrust dynamics, the control acceleration \(\mathbf{u}=[u_r,u_\theta,u_z]^T\) can be expressed as a function of \(\mathbf{q}\), \(\dot{\mathbf{q}}\), and \(\ddot{\mathbf{q}}\) as
\begin{equation}
	\mathbf{u}=
	\begin{bmatrix}
		\ddot{r}-r\dot{\theta}^2+\dfrac{\mu r}{s^3}\\
		2\dot{r}\dot{\theta}+r\ddot{\theta}\\
		\ddot{z}+\dfrac{\mu z}{s^3}\end{bmatrix}. \label{eq:flatness2}
\end{equation}
Thus, the control acceleration is a function of the cylindrical coordinates and their derivatives up to second order, \(\mathbf{u}\equiv\mathbf{u}(\mathbf{q},\dot{\mathbf{q}},\ddot{\mathbf{q}})\). Consequently, a sufficiently smooth trajectory \(\mathbf{q}(t)\) uniquely determines its associated control \(\mathbf{u}(t)\) through Eq.~\eqref{eq:flatness2}. A continuous control profile requires the trajectory to be at least \(C^2\)-continuous.

In the problem under consideration, the differential flatness property follows from expressing the fully actuated translational dynamics in cylindrical coordinates. By contrast, the complete vector of osculating orbital elements, including modified equinoctial elements, cannot be prescribed as an arbitrary trajectory and then directly inverted through the Gauss variational equations to obtain a unique control acceleration. Element-based shaping methods therefore require additional compatibility relations or a separate inverse mapping between the shaped variables and the physical trajectory.

\subsection{B-spline parameterization of cylindrical coordinates}

The spacecraft cylindrical coordinates
$\mathbf{q}=[r,\,\theta,\,z]^T$ are parameterized using a B-spline:
\begin{equation}
	\mathbf{q}(\tau)
	= \sum_{i=1}^{n_c} \mathbf{c}_{i} B_{i,p}(\tau),
	\quad
	\tau=\dfrac{t-t_0}{t_f-t_0},
	\label{eq:bspline}
\end{equation}
where \(\mathbf{c}_i\in\mathbb{R}^3\) denotes the \(i\)th control point, \(B_{i,p}\) is a piecewise polynomial basis function of degree \(p\), and \(n_c\) is the number of control points. The independent variable is the normalized time \(\tau=(t-t_0)/T\in[0,1]\), with \(T=t_f-t_0\).

For a B-spline of degree \(p\) with \(n_c\) control points, the number of knots is \(m=p+n_c+1\). The knot vector \(\boldsymbol{\Xi}=[\xi_1,\ldots,\xi_m]\) defines the time partition and continuity of the B-spline. A basis function \(B_{i,p}\) is nonzero only over the knot span \(\tau\in[\xi_i,\xi_{i+p+1})\). Therefore, its associated control point \(\mathbf{c}_i\) influences the summation in Eq.~\eqref{eq:bspline} only over that interval. At a simple interior knot, the B-spline is \(C^{p-1}\)-continuous. If an interior knot has multiplicity \(\nu\), the continuity decreases to \(C^{p-\nu}\). The basis functions \(B_{i,p}\) are defined using the Cox--de Boor recursion formula,
\begin{equation}
	\begin{aligned}
		B_{i,0}(\tau) &=
		\begin{cases}
			1, & \xi_i \le \tau < \xi_{i+1}, \\
			0, & \text{otherwise,}
		\end{cases}\\
		B_{i,p}(\tau) &=
		\dfrac{\tau - \xi_i}{\xi_{i+p}-\xi_i}\,B_{i,p-1}(\tau)
		+
		\dfrac{\xi_{i+p+1}-\tau}{\xi_{i+p+1}-\xi_{i+1}}\,B_{i+1,p-1}(\tau),
	\end{aligned}
	\label{eq:coxdeboorp}
\end{equation}
where a term with a zero denominator is defined to be zero. At the right endpoint, the standard clamped-spline convention \(B_{n_c,p}(1)=1\) is used. Equivalently, the basis values at \(\tau=1\) are defined by their left-hand limits. These conventions make the recursion well defined for the repeated endpoint knots of a clamped B-spline.

The inverse dynamics in Eq.~\eqref{eq:flatness2} require the first and second normalized-time derivatives, $\mathbf{q}'=[r',\,\theta',\,z']^T$ and $\mathbf{q}''=[r'',\,\theta'',\,z'']^T$:
\begin{equation}
	\begin{aligned}
		\mathbf{q}'(\tau) &=\sum_{i=1}^{n_c-1} \mathbf{d}_{i} B_{i+1,p-1}(\tau), &
		\mathbf{d}_i &= p\,\dfrac{\mathbf{c}_{i+1}-\mathbf{c}_i}{\xi_{i+p+1}-\xi_{i+1}}, \\
		\mathbf{q}''(\tau) &= \sum_{i=1}^{n_c-2} \mathbf{e}_{i} B_{i+2,p-2}(\tau), &
		\mathbf{e}_i &= (p-1)\,\dfrac{\mathbf{d}_{i+1}-\mathbf{d}_i}{\xi_{i+p+1}-\xi_{i+2}}.
	\end{aligned}
	\label{eq:bspline_derivatives}
\end{equation}
All basis functions in Eq.~\eqref{eq:bspline_derivatives} are evaluated on the knot vector \(\boldsymbol{\Xi}\). With $(\cdot)'=\dfrac{\mathrm{d}(\cdot)}{\mathrm{d}\tau}$, the required physical-time derivatives $\dot{\mathbf{q}}=[\dot{r},\,\dot{\theta},\,\dot{z}]^T$ and $\ddot{\mathbf{q}}=[\ddot{r},\,\ddot{\theta},\,\ddot{z}]^T$ follow as
\begin{equation}
	\dot{\mathbf{q}}(\tau)=\dfrac{\mathbf{q}'(\tau)}{T},
	\quad
	\ddot{\mathbf{q}}(\tau)=\dfrac{\mathbf{q}''(\tau)}{T^2}.
	\label{eq:physical_time_derivatives}
\end{equation}
Each differentiation reduces the degree of the basis functions by one, and the highest-order piecewise-nonzero derivative is the \(p\)th derivative. Because the inverse dynamics require a continuous second derivative, the present formulation uses \(p\geq3\) with simple interior knots. The B-spline and its derivatives can be evaluated efficiently using the same recursive basis-function algorithm.

\subsubsection{Compact B-spline notation}

To simplify the notation, the B-spline expansions in Eqs.~\eqref{eq:bspline} and \eqref{eq:bspline_derivatives} are expressed in matrix form. The \(i\)th control point is defined as
\begin{equation}
	\mathbf{c}_i =[
		c_{r,i},\,
		c_{\theta,i},\,
		c_{z,i}]^T,\quad i=1,\ldots,n_c.
\end{equation}
Then, the element-wise control points associated with each cylindrical coordinate $\{r,\,\theta,\,z\}$ are stacked in the following control-point vectors:
\begin{equation}
	\begin{aligned}
	\mathbf{c}_r =[c_{r,1},\hdots,c_{r,n_c}]^T,\\
	\mathbf{c}_\theta =[c_{\theta,1},\hdots,c_{\theta,n_c}]^T,\\
	\mathbf{c}_z =[c_{z,1},\hdots,c_{z,n_c}]^T.
	\end{aligned}
	\label{eq:controlpoint_vector}
\end{equation}
The following basis vectors stack the basis functions of degrees $p$, $p-1$, and $p-2$, respectively:
\begin{equation}
	\mathbf{b}_p(\tau)=
	\begin{bmatrix}
		B_{1,p}(\tau)\\
		\vdots\\
		B_{n_c,p}(\tau)
	\end{bmatrix},\quad\quad
	\mathbf{b}_{p-1}(\tau)=
	\begin{bmatrix}
		B_{2,p-1}(\tau)\\
		\vdots\\ 
		B_{n_c,p-1}(\tau)
	\end{bmatrix},\quad\quad 
	\mathbf{b}_{p-2}(\tau)=
	\begin{bmatrix}
		B_{3,p-2}(\tau)\\
		\vdots\\
		B_{n_c,p-2}(\tau)
	\end{bmatrix}.\label{eq:basis_vector}
\end{equation}

Using Eqs.~\eqref{eq:controlpoint_vector}--\eqref{eq:basis_vector}, the B-spline expansion of Eq.~\eqref{eq:bspline} can be compactly expressed as
\begin{equation}
	\mathbf{q}(\tau)=
	\begin{bmatrix}
		\mathbf{c}_r\cdot\mathbf{b}_p(\tau)\\
		\mathbf{c}_\theta\cdot\mathbf{b}_p(\tau)\\
		\mathbf{c}_z\cdot\mathbf{b}_p(\tau)
	\end{bmatrix}.\label{eq:position_expansion}
\end{equation}
Applying the same notation to the first and second physical-time derivatives in Eq.~\eqref{eq:physical_time_derivatives} gives
\begin{equation}
	\dot{\mathbf{q}}(\tau;\mathbf{C})=\dfrac{1}{T}
	\begin{bmatrix}
		\mathbf{d}_r\cdot\mathbf{b}_{p-1}(\tau)\\
		\mathbf{d}_\theta\cdot\mathbf{b}_{p-1}(\tau)\\
		\mathbf{d}_z\cdot\mathbf{b}_{p-1}(\tau)\\
	\end{bmatrix},\quad
	\ddot{\mathbf{q}}(\tau;\mathbf{C})=\dfrac{1}{T^2}
	\begin{bmatrix}
		\mathbf{e}_r\cdot\mathbf{b}_{p-2}(\tau)\\
		\mathbf{e}_\theta\cdot\mathbf{b}_{p-2}(\tau)\\
		\mathbf{e}_z\cdot \mathbf{b}_{p-2}(\tau)\\
	\end{bmatrix}.
	\label{eq:derivative_expansion}
\end{equation}
Here, the derivative control-point coefficients for each coordinate are stacked as
\begin{equation}
	\begin{aligned}
	\mathbf{d}_q=[
		d_{q,1},\hdots,d_{q,n_c-1}]^T,\\
		\mathbf{e}_q=[
		e_{q,1},\hdots,e_{q,n_c-2}]^T,
	\end{aligned}
\end{equation}
where $q\in\{r,\theta,z\}$ denotes the corresponding cylindrical coordinate.

\subsubsection{Clamped B-spline for analytical satisfaction of boundary conditions}

\begin{figure}[htbp!]
	\centering
	\includegraphics[width=0.8\linewidth, trim={0cm 0.25cm 0cm 0cm}, clip]{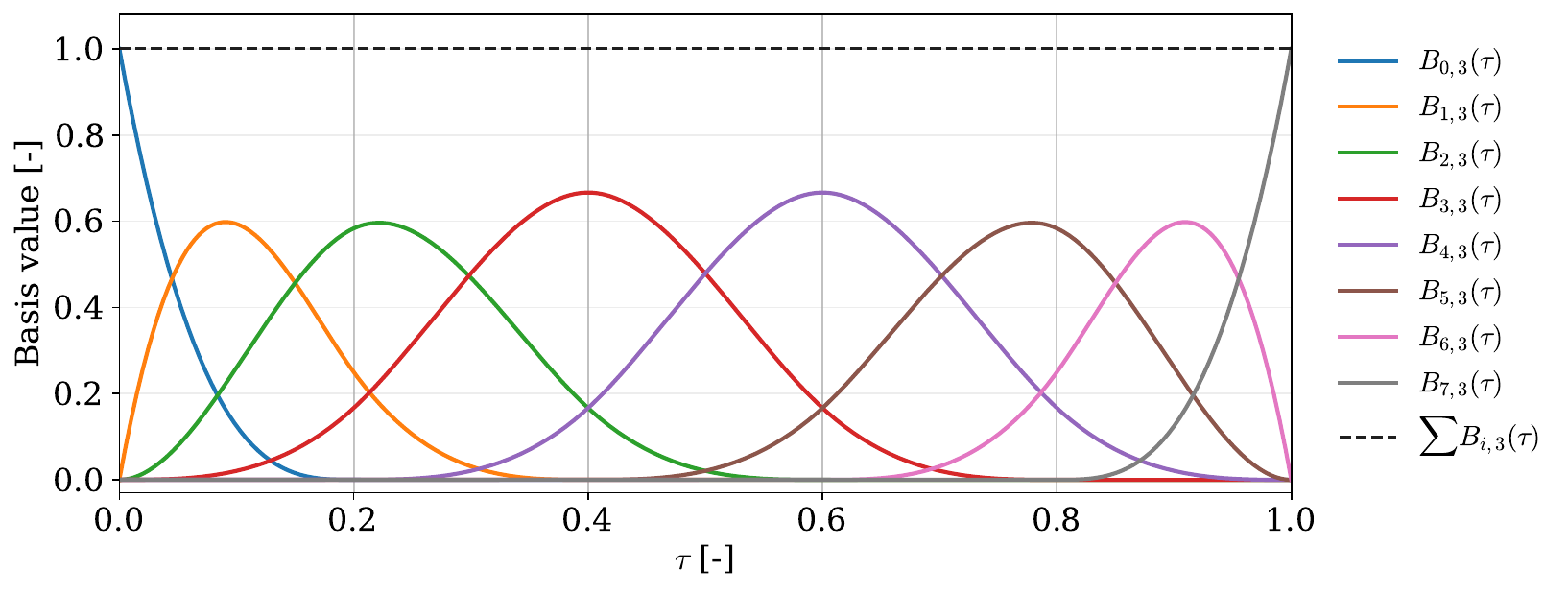}
	\caption{Basis functions for a clamped cubic B-spline.}
	\label{fig:bspline_basis}
\end{figure}

Repeating the knots at the endpoints, \(\tau=0\) and \(\tau=1\), permits boundary conditions to be prescribed by construction. In a clamped B-spline, each endpoint knot is repeated \(p+1\) times, producing the basis-function structure shown in Fig.~\ref{fig:bspline_basis}.

Using a clamped B-spline, the boundary conditions
\(\{\mathbf{q}_0,\dot{\mathbf{q}}_0\}\) and
\(\{\mathbf{q}_f,\dot{\mathbf{q}}_f\}\) can be prescribed analytically by fixing the following control points:
\begin{equation}
	\begin{aligned}
		\mathbf{c}_1 &= \mathbf{q}_0, &
		\mathbf{c}_{2} &= \mathbf{q}_0 + \dfrac{\xi_{p+2}}{p}\,T\dot{\mathbf{q}}_0,\\
		\mathbf{c}_{n_c} &= \mathbf{q}_f, &
		\mathbf{c}_{n_c-1} &= \mathbf{q}_f - \dfrac{1-\xi_{n_c}}{p}\,T\dot{\mathbf{q}}_f.
	\end{aligned}
	\label{eq:controlpoint_fixed}
\end{equation}
Here,
\begin{equation}
	\begin{aligned}
	\dot{\mathbf{q}}_0&=[v_{r_0},\,v_{\theta_0}/r_0,\,v_{z_0}]^T,\\
	\dot{\mathbf{q}}_f&=[v_{r_f},\,v_{\theta_f}/r_f,\,v_{z_f}]^T
	\end{aligned}
\end{equation}
are the physical-time derivatives of the cylindrical coordinates. In Eq.~\eqref{eq:controlpoint_fixed}, the factors \(T\) appear because the B-spline is differentiated with respect to the normalized time \(\tau=(t-t_0)/T\). The clamped B-spline therefore encodes the boundary conditions of problem~\eqref{prob:main} by construction.

After the first two and last two control points are prescribed by Eq.~\eqref{eq:controlpoint_fixed}, the remaining degrees of freedom are the \(n_c-4\) interior control points. These variables are stacked in the free-control-point vector $\mathbf{C}\in\mathbb{R}^{3(n_c-4)}$ as
\begin{equation}
\mathbf{C}=[\mathbf{c}^T_3,\hdots,\mathbf{c}^T_{n_c-2}]^T,
\end{equation}
which collects all decision variables.

\FloatBarrier
\subsection{Explicit B-spline expression for the control acceleration}

Combining the inverse dynamics in Eq.~\eqref{eq:flatness2} with the B-spline parameterization in Eq.~\eqref{eq:bspline} yields an explicit expression for the control acceleration $\mathbf{u}=[u_r,\,u_\theta,\,u_z]^T$ in terms of the B-spline basis functions and control points.

First, the cylindrical velocity $\mathbf{v}=[v_r,\,v_\theta,\,v_z]^T$ is expressed as a function of the B-spline variables. Substituting Eqs.~\eqref{eq:position_expansion}-\eqref{eq:derivative_expansion} into Eq.~\eqref{eq:flatness1} yields
\begin{equation}
	\mathbf{v}(\tau;\mathbf{C})=\dfrac{1}{T}
	\begin{bmatrix}
		\mathbf{d}_r\cdot\mathbf{b}_{p-1}\\
		(\mathbf{c}_r\cdot\mathbf{b}_p)\cdot\left(\mathbf{d}_\theta\cdot\mathbf{b}_{p-1}\right)\\
		\mathbf{d}_z\cdot\mathbf{b}_{p-1}\\
	\end{bmatrix}.
	\label{eq:velocity}
\end{equation}
The dependence of the basis vectors on normalized time is omitted for brevity. Substituting Eqs.~\eqref{eq:position_expansion}--\eqref{eq:derivative_expansion} into Eq.~\eqref{eq:flatness2} then gives
\begin{equation}
	\mathbf{u}(\tau;\mathbf{C}) =
	\dfrac{1}{T^2}
	\begin{bmatrix}
		\mathbf{e}_r\cdot\mathbf{b}_{p-2}-\left(\mathbf{c}_r\cdot\mathbf{b}_p\right)\cdot
		\left(\mathbf{d}_\theta\cdot\mathbf{b}_{p-1}\right)^2
		+\dfrac{T^2\mu\,\mathbf{b}_p\cdot\mathbf{c}_r}
		{\left[(\mathbf{c}_r\cdot\mathbf{b}_p)^2+
			(\mathbf{c}_z\cdot\mathbf{b}_p)^2\right]^{3/2}}\\
		2\left(\mathbf{d}_r\cdot\mathbf{b}_{p-1}\right)\cdot
		\left(\mathbf{d}_\theta\cdot\mathbf{b}_{p-1}\right)+\left(\mathbf{c}_r\cdot\mathbf{b}_p\right)\cdot\left(\mathbf{e}_\theta\cdot\mathbf{b}_{p-2}\right)\\
		\mathbf{e}_z\cdot\mathbf{b}_{p-2}
		+\dfrac{T^2 \mu\,\mathbf{c}_z\cdot\mathbf{b}_p}
		{\left[(\mathbf{c}_r\cdot\mathbf{b}_p)^2+
			(\mathbf{c}_z\cdot\mathbf{b}_p)^2\right]^{3/2}}
	\end{bmatrix}.
	\label{eq:control_compact}
\end{equation}
Equations~\eqref{eq:velocity}--\eqref{eq:control_compact} depend on the free-control-point vector $\mathbf{C}$ because $\mathbf{c}_q$, $\mathbf{d}_q$, and $\mathbf{e}_q$, for $q\in\{r,\theta,z\}$, are affine functions of $\mathbf{C}$ after the boundary conditions in Eq.~\eqref{eq:controlpoint_fixed} are imposed. Thus, satisfying the dynamics and boundary conditions by construction results in more complex, but tractable, control and cost expressions.

\subsection{Finite-dimensional nonlinear programming formulation}

The $\Delta V$ cost of the continuous optimization problem~\eqref{prob:main} can now be expressed as a function of the free-control-point vector $\mathbf{C}$:
\begin{equation}
		\Delta V (\mathbf{C})=T\int^1_0\sqrt{u^2_r(\tau;\mathbf{C})+u^2_\theta(\tau;\mathbf{C})+u^2_z(\tau;\mathbf{C})}\,\mathrm{d}\tau,\label{eq:dv}
\end{equation}
where Eq.~\eqref{eq:control_compact} is used implicitly. The full expression is omitted for brevity.

Because the dynamics are satisfied through the flatness mapping in Eq.~\eqref{eq:flatness2} and the boundary conditions are prescribed analytically by Eq.~\eqref{eq:controlpoint_fixed}, the continuous optimization problem~\eqref{prob:main} is reduced to the finite-dimensional nonlinear program
\begin{mini}
	{\mathbf{C}}
	{\Delta V(\mathbf{C}) = T \int_{0}^{1} \left\lVert \mathbf{u}(\tau;\mathbf{C}) \right\rVert \mathrm{d}\tau,}
	{\label{prob:staticprogram}}{}
	\addConstraint{\mathbf{C}_{\text{min}}\leq\mathbf{C}\leq\mathbf{C}_{\text{max}},}
\end{mini}
The number of decision variables is \(3(n_c-4)\), the dimension of the B-spline free-control-point vector \(\mathbf{C}\). Bounds are imposed on the control points to restrict the search to a physically meaningful domain:
\begin{equation}
	\begin{aligned}
	\mathbf{C}_{\text{min}}&=[
		\mathbf{c}^T_{\text{min}},\hdots,
		\mathbf{c}^T_{\text{min}}]^T,\\
	\mathbf{C}_{\text{max}}&=[
		\mathbf{c}^T_{\text{max}},\hdots,
		\mathbf{c}^T_{\text{max}}]^T,
	\end{aligned}
\end{equation}
where \(\mathbf{c}_{\text{min}}=[c_{r\text{min}},\,c_{\theta\text{min}},\,c_{z\text{min}}]^T\) and \(\mathbf{c}_{\text{max}}=[c_{r\text{max}},\,c_{\theta\text{max}},\,c_{z\text{max}}]^T\). In particular, the flatness mapping requires \(r(t)>0\). Because the B-spline basis functions are nonnegative and form a partition of unity, imposing the strictly positive radial control-point bound \(c_{r,i}\geq c_{r\text{min}}>0\) guarantees \(r(t)\geq c_{r\text{min}}>0\) throughout the transfer.

Note that the transformation of the continuous optimization problem~\eqref{prob:main} into the finite static program~\eqref{prob:staticprogram} is subject to a convergence--optimality tradeoff that is inherent to shape-based methods. Solving the nonlinear program~\eqref{prob:staticprogram} yields a locally optimal candidate within the selected finite-dimensional B-spline shape space. This candidate may be suboptimal with respect to the original continuous problem~\eqref{prob:main}, and the optimality gap depends on both the representational capacity of the clamped B-spline and convergence to a favorable local solution. However, the static nonlinear program~\eqref{prob:staticprogram} encodes the dynamics and boundary conditions by construction, making it convenient for obtaining fast solutions to a complex optimization problem.

\section{Numerical implementation}

This section details the numerical implementation of the proposed B-spline shaping method. To benchmark its performance, the hodographic shaping approach of \citep{Gondelach2015} is also implemented and described.

\subsection{B-spline shaping implementation}

The B-spline shaping nonlinear program~\eqref{prob:staticprogram} is formulated symbolically using CasADi~\citep{Andersson2019} and solved with IPOPT~\citep{Wachter2006}. The implementation is written in Python. IPOPT requires an initial estimate of the free B-spline coefficients for its interior-point algorithm. The estimate used here is based on a quintic polynomial profile in modified equinoctial elements (MEE)~\citep{Walker1985}:
\begin{equation}
	\begin{aligned}
		\pmb{\alpha}(\tau)
		&=
		\pmb{\alpha}_0
		+\lambda(\tau)(\pmb{\alpha}_f-\pmb{\alpha}_0),\\
		\lambda(\tau)&=10\tau^3-15\tau^4+6\tau^5,
	\end{aligned}\label{eq:initguess}
\end{equation}
where \(\pmb{\alpha}=[p,\,f,\,g,\,h,\,k,\,L]^T\) contains the MEE. The final true longitude has to include the number of additional revolutions given by $N$. Equation~\eqref{eq:initguess} gives zero first and second MEE derivatives at both boundaries:
\(\pmb{\alpha}'(0)=\pmb{\alpha}''(0)=\pmb{\alpha}'(1)=\pmb{\alpha}''(1)=0\). The resulting MEE profile is converted to cylindrical coordinates, and the free B-spline control points \(\mathbf{C}\) are fitted to that profile in a least-squares sense while satisfying Eq.~\eqref{eq:controlpoint_fixed}. The initial control polygon therefore approximates the quintic MEE profile while remaining consistent with the endpoint conditions.

An important aspect of the B-spline implementation is the computation of the \(\Delta V\) cost integral in Eq.~\eqref{prob:staticprogram}. This cost is challenging to optimize because its gradient is undefined at \(\mathbf{u}=\mathbf{0}\). In numerical experiments, the B-spline parameterization exhibited formal convergence difficulties near these points. To improve convergence, the integrand is smoothed around \(\mathbf{u}=\mathbf{0}\) as
\begin{equation}
	\Delta V \approx T \int_{0}^{1} \left( \sqrt{\lVert\mathbf{u}(\tau,\mathbf{C})\rVert^2+\epsilon^2}-\epsilon \right) \mathrm{d}\tau,
\end{equation}
where \(\epsilon=10^{-6}\) in canonical acceleration units. When reporting the $\Delta V$ cost, the smoothing is deactivated and the original integral in problem~\eqref{prob:staticprogram} is evaluated. Composite Gauss--Legendre quadrature is used over the nonzero B-spline knot spans, with six Gauss points per span.

For the B-splines used in this work, the distinct interior knots are equally spaced over \(\tau\in[0,1]\). The following control-point bounds are imposed in the transfers analyzed in Section~\ref{sec5}:
\begin{equation}
\begin{aligned}
10^{-4}\>\text{AU}&\leq c_{r,i}\leq 20\>\text{AU},\\
\theta_0-2\pi &\leq c_{\theta,i} \leq \theta_f+2\pi (N+1),\\
-20\>\text{AU}&\leq c_{z,i}\leq 20\>\text{AU}.\\
\end{aligned}
\end{equation}
The small positive lower bound on the radial control points prevents the trajectory from approaching the singularity at \(r=0\). The upper radial and axial bounds provide a wide search margin. The angular bounds cover the prescribed initial and unwrapped final angles with one additional revolution on each side. The B-spline implementation uses the heliocentric canonical units
\begin{equation}
	\mathrm{DU}=1~\mathrm{AU},\quad \mathrm{TU}=\frac{1}{2\pi}~\mathrm{yr},\quad \mathrm{VU}=2\pi~\dfrac{\mathrm{AU}}{\mathrm{yr}},\quad \mathrm{A}_U=4\pi^2~\dfrac{\mathrm{AU}}{\mathrm{yr}^2},\quad \mu_\odot^\ast=1.
\end{equation}

\subsection{Hodographic shaping benchmark}\label{sec:42}

The hodographic shaping method of \citep{Gondelach2015} is implemented as a benchmark. Whereas the proposed method parameterizes the cylindrical coordinates, hodographic shaping parameterizes the cylindrical velocity \(\mathbf{v}=[v_r,\,v_\theta,\,v_z]^T\) using analytical basis functions, including constants, powers, trigonometric terms, and their combinations. In the hodographic formulation, the three position boundary conditions are expressed as integral constraints:
\begin{equation}
	\begin{aligned}
		T\int_{0}^{1} v_r\,\mathrm{d}\tau
		&= r_f-r_0,\\
		T\int_{0}^{1}\frac{v_\theta}{r}\,\mathrm{d}\tau
		&= \theta_f-\theta_0+2\pi N,\\
		T\int_{0}^{1} v_z\,\mathrm{d}\tau
		&= z_f-z_0.
	\end{aligned}
	\label{eq:hodographic_integral_constraints}
\end{equation}
Together with the six endpoint velocity conditions, these relations give nine scalar boundary conditions. A low-order hodographic solution uses nine coefficients, which are fully determined by those conditions and require no optimization. Additional basis functions introduce free coefficients and produce a high-order solution. This work compares the B-spline method with the best high-order hodographic configurations reported on the target-specific Pareto fronts in Figs.~5, 8, 11, and 14 of \cite{Gondelach2015}. Table~\ref{tab:gondelach_basis} summarizes these basis functions, where \(\tau=(t-t_0)/(t_f-t_0)\) and \(\omega_N=2\pi(N+0.5)\tau\).

\begin{table}[h!]
	\center
	\caption{Best high-order hodographic shaping basis functions found by \cite{Gondelach2015}.}
	\label{tab:gondelach_basis}
	\setlength{\tabcolsep}{3pt}
	\begin{tabular*}{\textwidth}{@{\extracolsep{\fill}}lll@{}}
		\toprule
		Target &
		Basis functions for $v_r$ and $v_\theta$ &
		Basis functions for $v_z$ \\
		\midrule
		Mars &
		$\{1,\tau,\tau^2,\tau\sin(\pi\tau),\tau\cos(\pi\tau)\}$ &
		$\{\cos\omega_N,\tau^3\cos\omega_N,\tau^3\sin\omega_N,\tau^4\cos\omega_N,\tau^4\sin\omega_N\}$ \\
		
		Mercury &
		$\{1,\tau,\tau^2,\tau\sin(2\pi\tau),\tau\cos(2\pi\tau)\}$ &
		$\{\cos(\pi\tau),\tau^5,\tau^6,\tau^6\cos(\pi\tau),\tau^6\sin(\pi\tau)\}$ \\
		
		1989 ML &
		$\{1,\tau,\tau^2,\tau\sin(2\pi\tau),\tau\cos(2\pi\tau)\}$ &
		$\{\cos(\pi\tau),\tau\cos(\pi\tau),\tau\sin(\pi\tau),\tau^6\cos(\pi\tau),\tau^6\sin(\pi\tau)\}$ \\
		
		Tempel 1 &
		$\{1,\tau,\sin(\pi\tau),\tau\sin(2\pi\tau),\tau\cos(2\pi\tau)\}$ &
		$\{\cos(\pi\tau),\tau^2,\tau^3,\tau^3\cos(\pi\tau),\tau^3\sin(\pi\tau)\}$ \\
		\bottomrule
	\end{tabular*}
\end{table}

Table~\ref{tab:gondelach_basis} contains 15 basis functions for each
target body. Because nine coefficients are determined by the boundary
conditions, six free coefficients remain in every case. In the present
implementation, the resulting unconstrained six-dimensional problem is
solved using the Nelder--Mead method provided by \texttt{SciPy} \cite{SciPy2020}. The
initial simplex is constructed using coordinate perturbations of
$10^{-2}$. A maximum of 1{,}000 objective-function evaluations is
allowed, with absolute tolerances of $10^{-6}$ for both the design
variables and the objective function.

At the first grid point, each value of $N$ is initialized using zero
free coefficients. Subsequent problems are warm-started using the
usable solution at the preceding departure date for the same $N$. If
that solution is unavailable, the usable solution at the preceding
time of flight is used. This continuation strategy reduces the
computational cost and improves convergence across the departure-date
and time-of-flight grid. The original implementation of
\cite{Gondelach2015} employed O'Neill's implementation of the
Nelder--Mead algorithm. The present implementation retains the same
derivative-free simplex formulation but uses the \texttt{SciPy}
	implementation. During optimization, each trajectory is sampled at 51 uniformly spaced
	values of normalized time. Analytical basis-function derivatives and
	analytical radial and axial integrals are used. Composite Simpson
	integration is used for the transverse angular constraint and the
	$\Delta V$ objective. For the comparative results, the $\Delta V$ integrals obtained with both
	hodographic and B-spline shaping are subsequently reevaluated in
	postprocessing using the same higher-accuracy composite Gauss--Legendre
	quadrature, so that the reported costs are evaluated on a consistent
	numerical basis.

\section{Numerical results and discussion}
\label{sec5}

This section presents the numerical results obtained using the proposed
B-spline shaping method described in Section~\ref{sec:3}. The method is
benchmarked against the high-order hodographic shaping approach described
in Section~\ref{sec:42}. Four low-thrust rendezvous scenarios from Earth
to different celestial bodies are considered. The computational efficiency
and solution quality of the two methods are then discussed. Finally, the
influence of the B-spline hyperparameters on the resulting control profiles
is examined.

\subsection{Low-thrust rendezvous with different targets}

The purpose of this subsection is to assess the performance of the proposed
B-spline approach for transfers to different celestial bodies. Following
\cite{Gondelach2015}, Mars, Mercury, the near-Earth asteroid 1989 ML, and
comet Tempel~1 are considered. All transfers depart from Earth. The
heliocentric osculating orbital elements of the target bodies at 00:00 UTC
on 1 January 2028 are listed in Table~\ref{tab:target_elements}. The elements
were obtained from the DE440 planetary ephemeris and the corresponding
small-body SPICE kernels using the SPICE Toolkit and its Python interface,
SpiceyPy~\cite{Acton1996,Annex2020}, and are expressed in the ECLIPJ2000
frame.

\begin{table}[htb!]
	\caption{Heliocentric osculating orbital elements of the target bodies
		at 00:00 UTC on 1 January 2028.}
	\label{tab:target_elements}
	\centering
	\begin{tabular}{@{}lrrrrrr@{}}
		\toprule
		Target & \(a\) [AU] & \(e\) [-] & \(i\) [$^{\circ}$] &
		\(\Omega\) [$^{\circ}$] & \(\omega\) [$^{\circ}$] &
		\(M\) [$^{\circ}$] \\
		\midrule
		Mars     & 1.52367 & 0.09353 &  1.84747 &  49.47202 & 286.75688 & 338.24330 \\
		Mercury  & 0.38710 & 0.20563 &  7.00331 &  48.29535 &  29.20320 & 265.05991 \\
		1989 ML  & 1.27274 & 0.13679 &  4.37885 & 104.26705 & 183.67239 & 290.68494 \\
		Tempel 1 & 3.30229 & 0.46352 & 10.45600 &  66.57384 & 184.80218 & 353.03934 \\
		\bottomrule
	\end{tabular}
\end{table}

\FloatBarrier
\subsubsection{Mars}

For the Earth--Mars case, transfer times ranging from 500 to
2{,}000 days and departure dates from 1 January 2028 to 30 December
2035 are considered. Both grids use a 20-day spacing. At each
departure-date and transfer-time combination, the number of spacecraft
revolutions is varied as $N\in\{0,1,2,3,4,5\}$, and the solution with
the lowest $\Delta V$ is retained.

\begin{figure}[htbp!]
	\centering
	\includegraphics[width=1.0\linewidth]
	{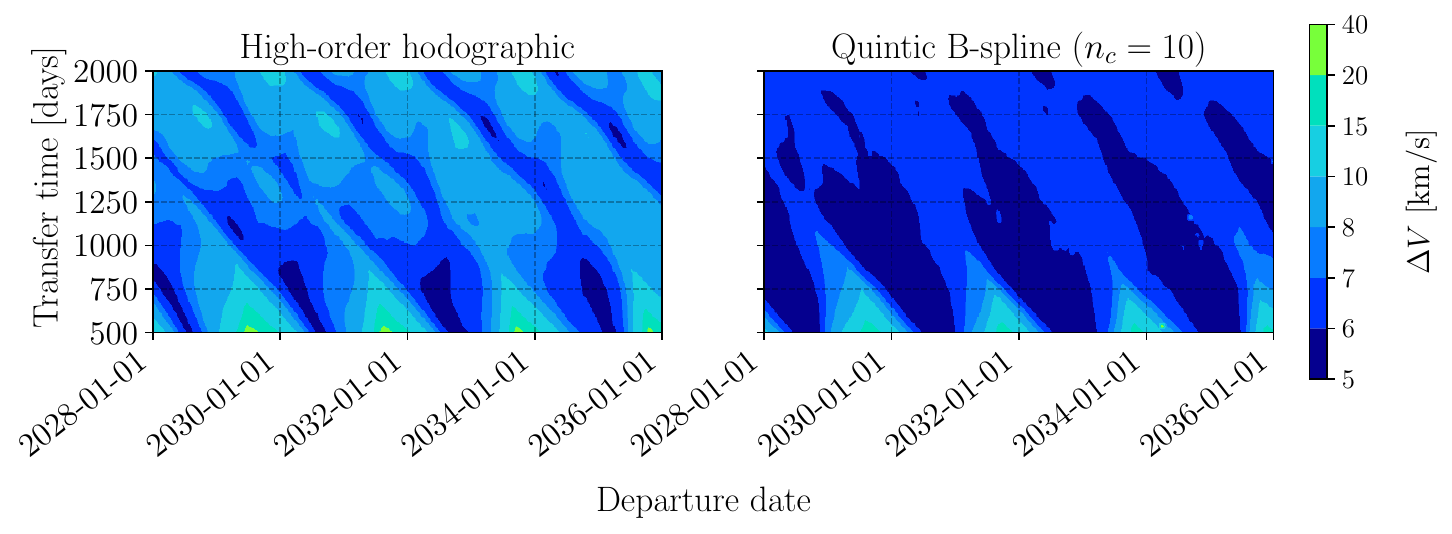}
	\caption{Comparison of Earth--Mars low-thrust rendezvous
		$\Delta V$ porkchop plots.}
	\label{fig:porkchopMars}
\end{figure}

Figure~\ref{fig:porkchopMars} shows the corresponding porkchop plots
for the high-order hodographic method and the quintic B-spline
configuration with 10 control points. The B-spline method produces
lower $\Delta V$ values over most of the grid, with a large proportion
of the solutions lying between 5 and 7~km/s. The high-order
hodographic porkchop exhibits a structure similar to that reported in
Figure~3 of \cite{Gondelach2015}, although exact agreement is not
expected because of differences in the departure window and
ephemerides. The periodic structure associated with the Earth--Mars
synodic period of approximately 2.13 years is also evident, with nearly
four repeated structures appearing over the eight-year departure
window.

\begin{figure}[htbp!]
	\centering
	\includegraphics[width=0.9\linewidth]
	{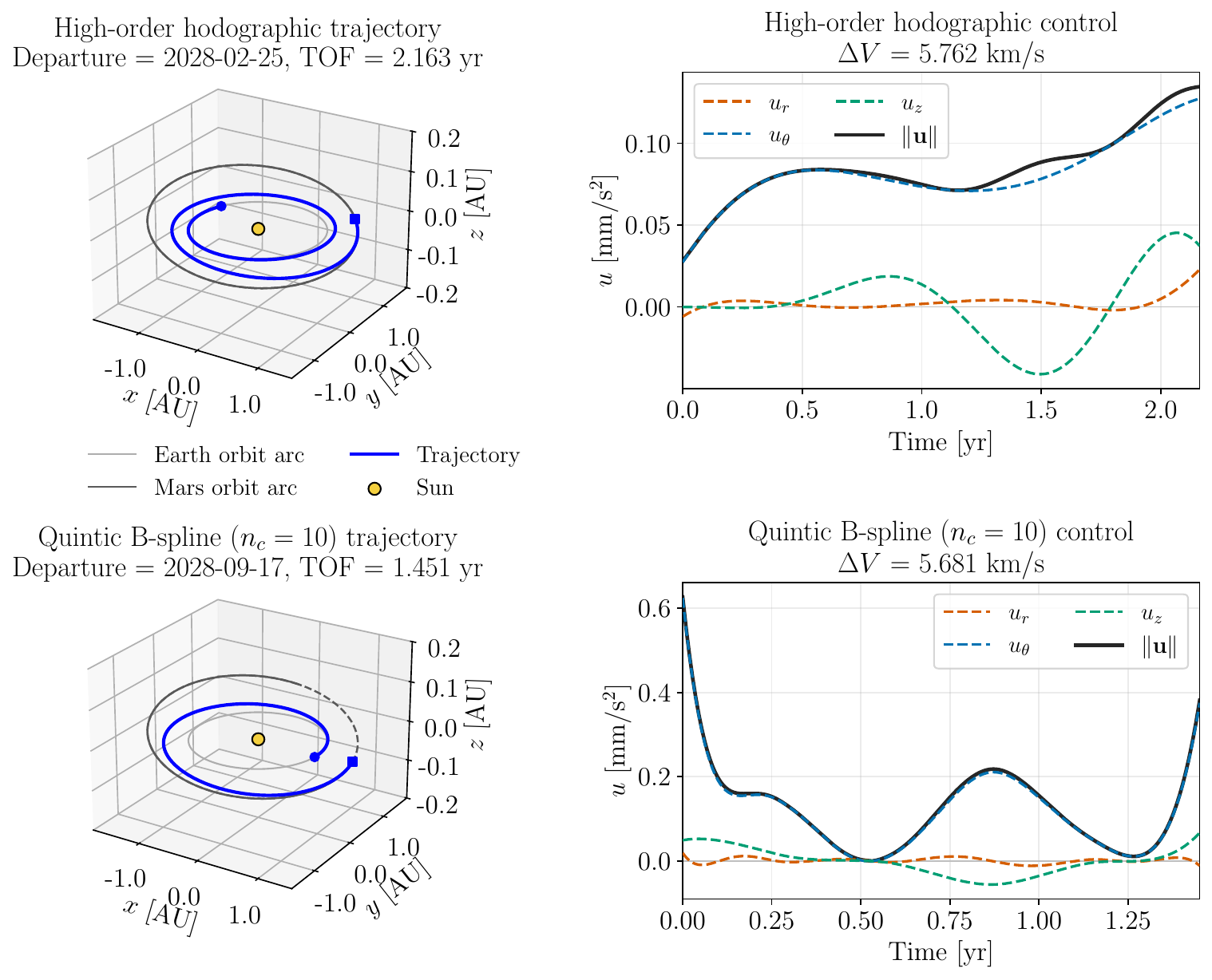}
	\caption{Lowest-$\Delta V$ Earth--Mars trajectories and control
		accelerations obtained using high-order hodographic shaping
		and a quintic B-spline with $n_c=10$.}
	\label{fig:Marsbest_comparison}
\end{figure}

To examine one Earth--Mars synodic period in greater detail, the
departure window is restricted to the interval from 1 January 2028 to
9 February 2030. Departure dates are sampled at 5-day intervals,
whereas transfer times from 500 to 2{,}000 days are sampled at 10-day
intervals. Four B-spline configurations are considered, corresponding
to cubic and quintic B-splines with 10 and 40 control points. Before considering the aggregate results, Figure~\ref{fig:Marsbest_comparison}
compares the lowest-$\Delta V$ trajectory found using high-order
hodographic shaping with that obtained using the quintic B-spline
configuration with 10 control points. The corresponding costs are
5.762 and 5.681~km/s, respectively. The B-spline solution therefore
reduces $\Delta V$ by approximately 1.4\% relative to
the hodographic solution. The two solutions occur at different
departure dates and transfer times: the hodographic transfer departs
on 25 February 2028 and lasts 790 days, whereas the B-spline transfer
departs on 17 September 2028 and lasts 530 days. Both solutions
correspond to $N=1$ and complete approximately one heliocentric
revolution before rendezvous with Mars. The thrust acceleration is predominantly directed along the positive
tangential direction, $u_\theta$, as expected for heliocentric orbit
raising. In the hodographic solution, the acceleration magnitude
generally increases toward arrival. The B-spline solution instead exhibits a high
initial acceleration followed by a nonmonotonic profile, with two
intermediate minima and a final increase near arrival.

\begin{figure}[htbp!]
	\centering
	\includegraphics[width=0.6\linewidth]
	{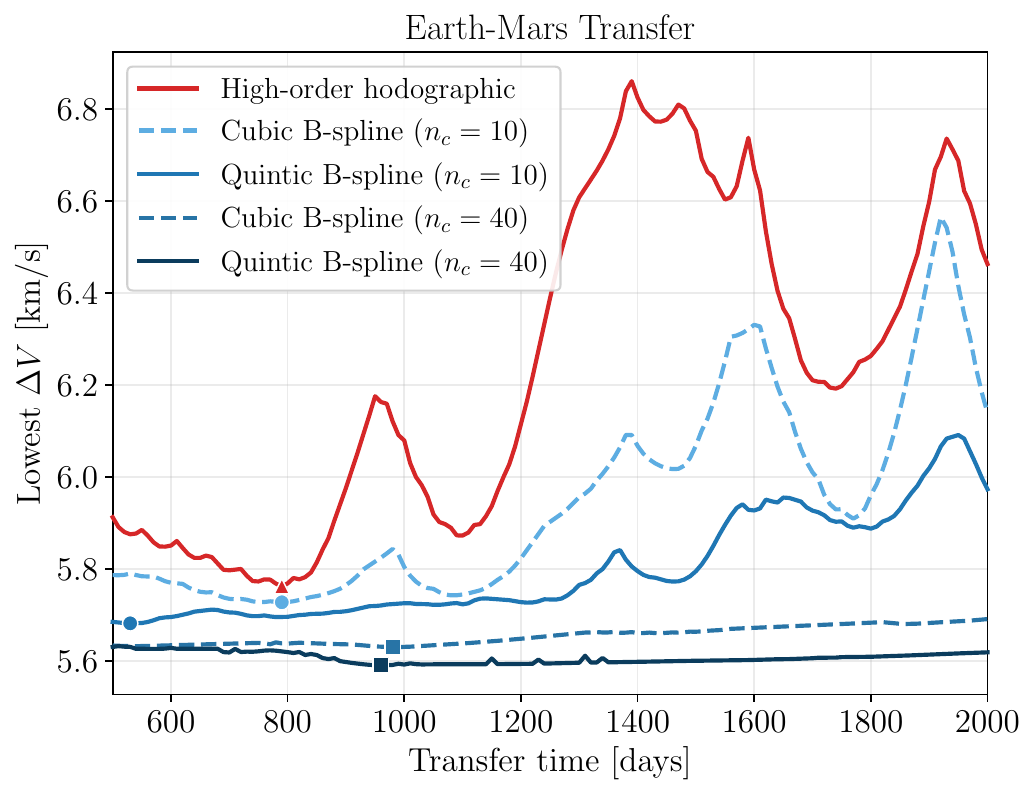}
	\caption{Minimum $\Delta V$ over the departure-date grid for each
		Earth--Mars transfer time from 1 January 2028 to 9 February 2030.}
	\label{fig:tofvsfuelMars}
\end{figure}

\begin{figure}[htbp!]
	\centering
	\includegraphics[width=0.6\linewidth]
	{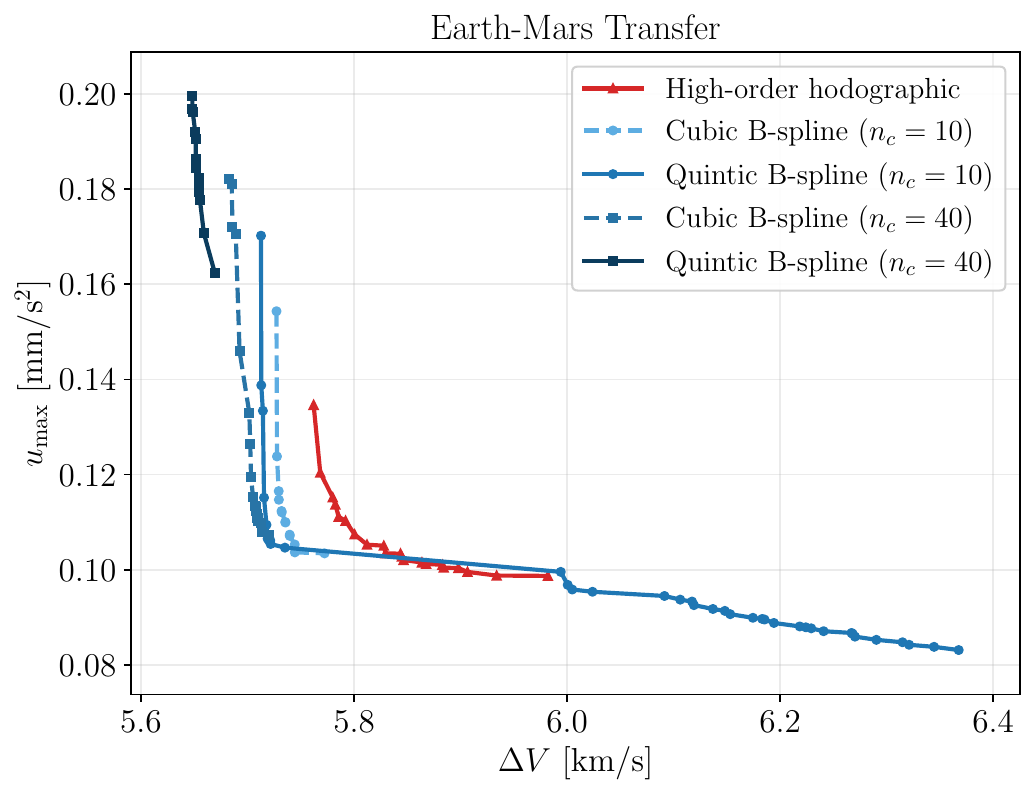}
	\caption{Pareto front of maximum control acceleration versus
		$\Delta V$ for Earth--Mars transfers from 1 January 2028 to
		9 February 2030.}
	\label{fig:paretoMars}
\end{figure}

The aggregate results over the refined grid are presented in
Figures~\ref{fig:tofvsfuelMars} and~\ref{fig:paretoMars}.
Figure~\ref{fig:tofvsfuelMars} shows the minimum $\Delta V$ obtained
over all departure dates for each transfer time. On the tested refined
grid and among the finite solutions retained, all four B-spline
configurations yield lower minimum $\Delta V$ values than the high-order
hodographic method over the considered range. In particular, the configurations with 40 control
points maintain $\Delta V$ values between approximately 5.6 and
5.7~km/s over nearly the entire transfer-time interval, with the
quintic configuration providing the lowest overall values.

Following \cite{Gondelach2015}, Figure~\ref{fig:paretoMars} shows the
Pareto front between the maximum control acceleration norm, $u_{\max}$, and
$\Delta V$. Lower values of $\Delta V$ are generally associated with
higher maximum accelerations, illustrating the trade-off between
propellant consumption and thrust capability. This comparison may
therefore support the preliminary selection of a transfer when an
upper limit on $u_{\max}$ is considered. However, $u_{\max}$ is neither
imposed as a constraint nor included in the optimization objective.
The resulting front should consequently be interpreted as an
a posteriori comparison of the computed solutions rather than as the
solution of a formally constrained multiobjective problem.

\FloatBarrier
\subsubsection{Mercury}

For the Earth--Mercury case, transfer times ranging from 100 to
1{,}400 days and departure dates from 1 January 2028 to 20 December
2033 are considered. Both grids use a 20-day spacing. The same
selection procedure used for Mars is applied, with the solution having
the lowest finite $\Delta V$ retained at each departure-date and
transfer-time combination. The revolution branches are
$N\in\{0,1,2,3,4,5\}$.

\begin{figure}[htbp!]
	\centering
	\includegraphics[width=1.0\linewidth]
	{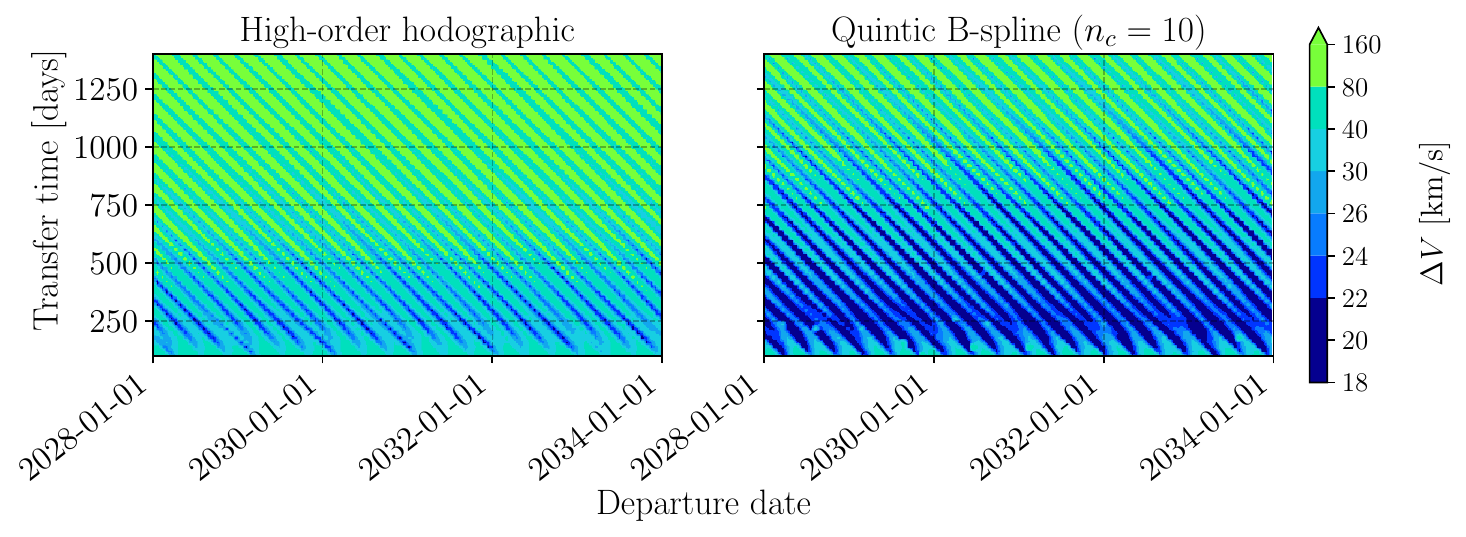}
	\caption{Comparison of Earth--Mercury low-thrust rendezvous
		$\Delta V$ porkchop plots.}
	\label{fig:porkchopMercury}
\end{figure}

Figure~\ref{fig:porkchopMercury} shows the corresponding porkchop plots
for the high-order hodographic method and the quintic B-spline
configuration with 10 control points. The B-spline method produces
substantially lower $\Delta V$ values over a large portion of the grid,
particularly at shorter transfer times. The diagonal bands in both
plots result from the Earth--Mercury synodic period of approximately
116 days, which is considerably shorter than that of Earth and Mars.

\begin{figure}[htbp!]
	\centering
	\includegraphics[width=0.9\linewidth]
	{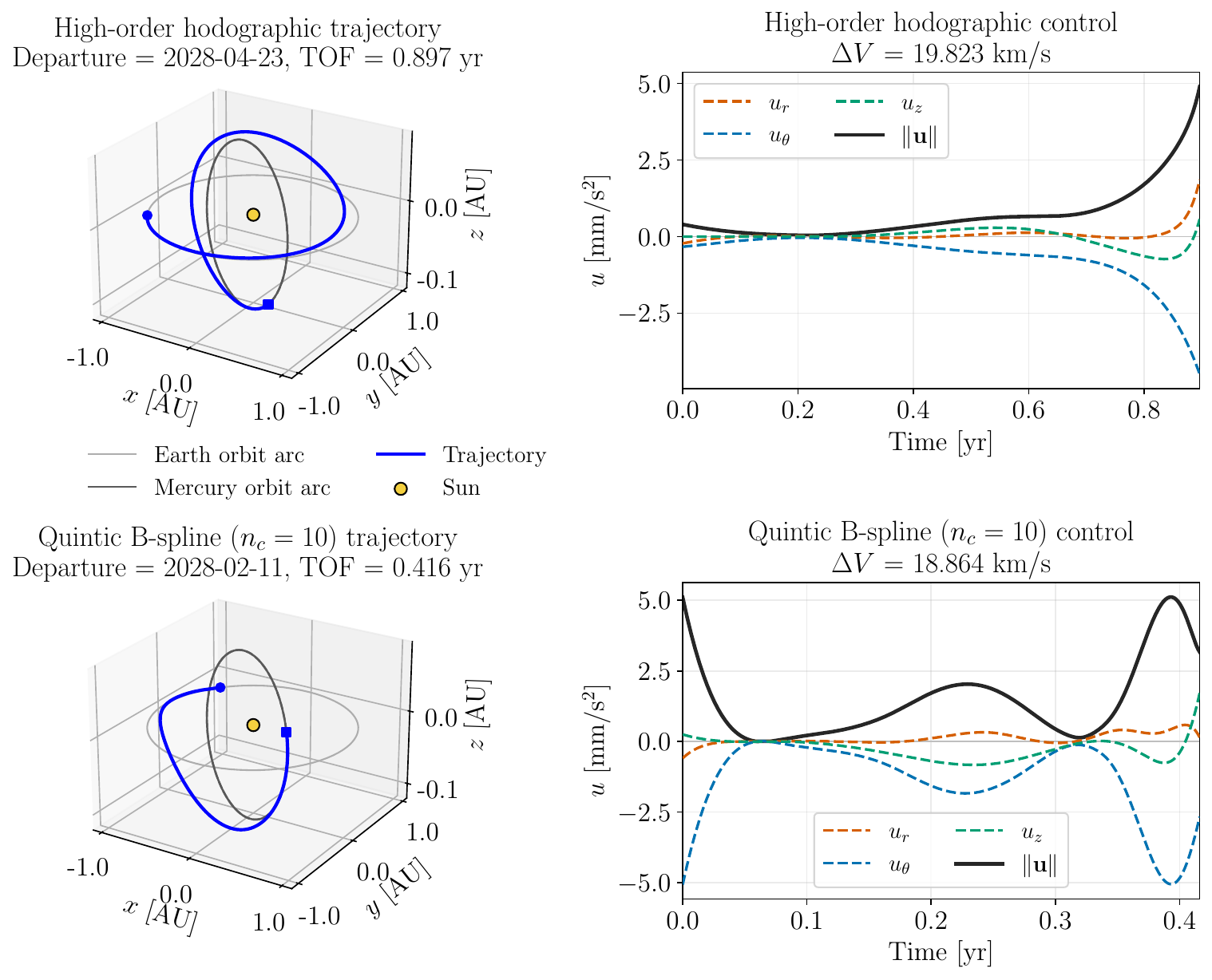}
	\caption{Lowest-$\Delta V$ Earth--Mercury trajectories and control
		accelerations obtained using high-order hodographic shaping
		and a quintic B-spline with $n_c=10$.}
	\label{fig:Mercurybest_comparison}
\end{figure}

For the refined analysis, the departure window is restricted to the
interval from 1 January to 26 April 2028, which covers approximately
one Earth--Mercury synodic period. Departure dates are sampled at
1-day intervals, whereas transfer times from 100 to 1{,}400 days are
sampled at 6.5-day intervals. The same six revolution branches,
\(N\in\{0,1,2,3,4,5\}\), are evaluated. As in the Mars analysis, cubic and
quintic B-splines with 10 and 40 control points are considered. Figure~\ref{fig:Mercurybest_comparison} compares the lowest-$\Delta V$
trajectory found using high-order hodographic shaping with that
obtained using the quintic B-spline configuration with 10 control
points. The corresponding costs are 19.823 and 18.864~km/s,
respectively. The B-spline solution therefore reduces $\Delta V$ by
approximately 4.8\% relative to the hodographic
solution. The hodographic transfer departs on 23 April 2028 and lasts
327.5 days, whereas the B-spline transfer departs on 11 February 2028
and lasts 152 days. The solutions also follow different revolution
branches, with $N=1$ for the hodographic trajectory and $N=0$ for the
B-spline trajectory. In both cases, the control acceleration is primarily directed along
the negative tangential direction, consistent with the reduction in
heliocentric orbital energy required to reach Mercury. The
hodographic control magnitude remains relatively low during the first
part of the transfer and increases sharply near arrival. The B-spline
solution exhibits a strongly nonmonotonic profile, with high
accelerations near departure and arrival, an intermediate thrust peak,
and two intervals of nearly vanishing control.

\begin{figure}[htbp!]
	\centering
	\includegraphics[width=0.6\linewidth]
	{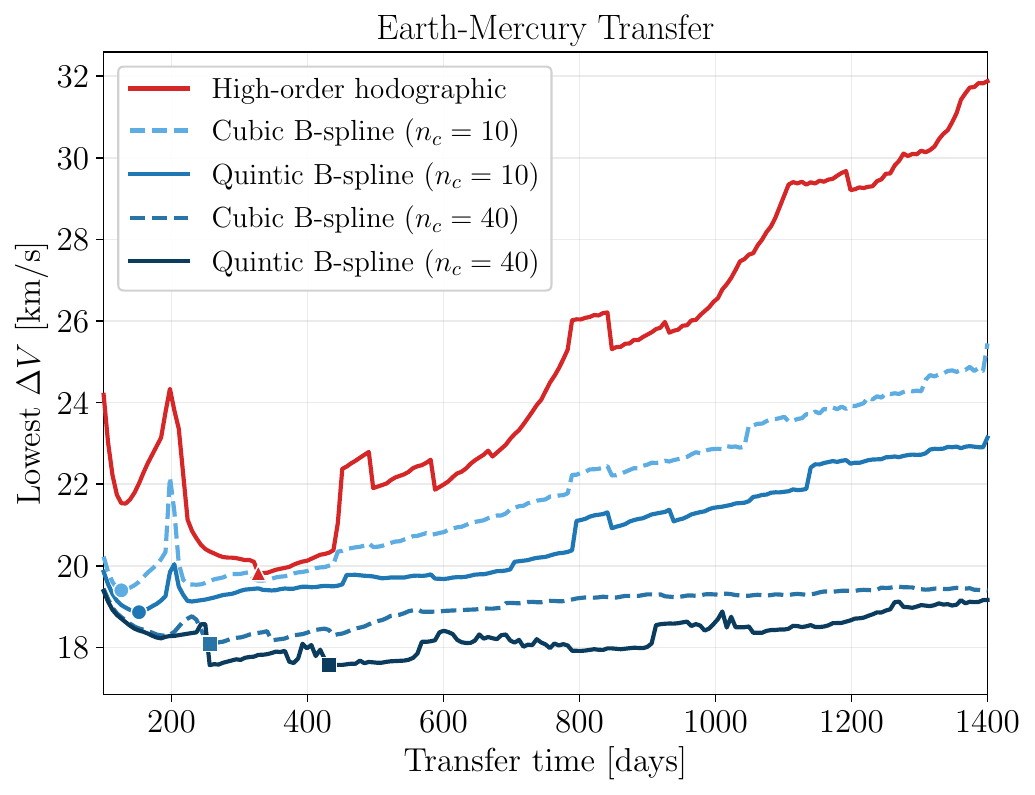}
	\caption{Minimum $\Delta V$ over the departure-date grid for each
		Earth--Mercury transfer time from 1 January to 26 April 2028.}
	\label{fig:tofvsfuelMercury}
\end{figure}

\begin{figure}[htbp!]
	\centering
	\includegraphics[width=0.6\linewidth]
	{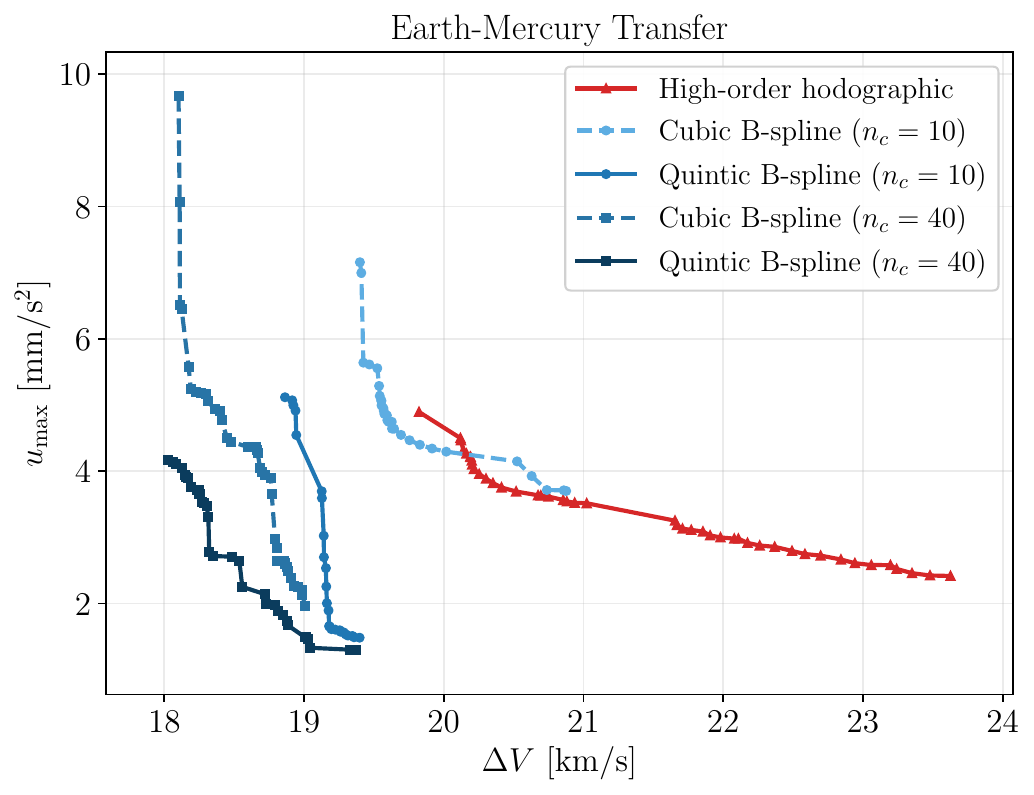}
	\caption{Pareto front of maximum control acceleration versus
		$\Delta V$ for Earth--Mercury transfers from 1 January to
		26 April 2028.}
	\label{fig:paretoMercury}
\end{figure}

The aggregate results over the refined grid are presented in
Figures~\ref{fig:tofvsfuelMercury} and~\ref{fig:paretoMercury}.
Figure~\ref{fig:tofvsfuelMercury} shows the minimum $\Delta V$ obtained
over all departure dates for each transfer time. The improvement
provided by B-spline shaping is particularly pronounced at short and
long transfer times. The configurations with 40 control points
maintain their minimum $\Delta V$ values between approximately 17.5
and 19.5~km/s throughout the transfer-time interval, whereas the
hodographic cost increases to approximately 32~km/s for the longest
transfers.

Figure~\ref{fig:paretoMercury} shows that the differences between the
methods are more pronounced than in the Earth--Mars case. In
particular, the quintic B-spline with 40 control points occupies the
lower-left region of the plot, combining comparatively low $\Delta V$
with low maximum control acceleration. The Earth--Mercury transfers
nevertheless require substantially higher maximum accelerations than
the Earth--Mars transfers. This increase is consistent with the larger
change in heliocentric orbital energy, the shorter transfer
opportunities, and the stronger solar gravitational field closer to
Mercury. As before, the Pareto curves represent an a posteriori
comparison because $u_{\max}$ is not explicitly constrained during
optimization.

\FloatBarrier
\subsubsection{Asteroid 1989 ML}

The third target considered is the near-Earth asteroid 1989 ML. Because its orbit is relatively close to that of Earth, orbital phasing is expected to play an important role in the transfer. Departure dates from 1 January 2028 to 30 December 2035 and transfer times from 100 to 1{,}000 days are considered, with both grids sampled at 20-day intervals. For this target, the number of spacecraft revolutions is restricted to $N\in\{0,1,2\}$.

\begin{figure}[htbp!]
	\centering
	\includegraphics[width=0.6\linewidth]
	{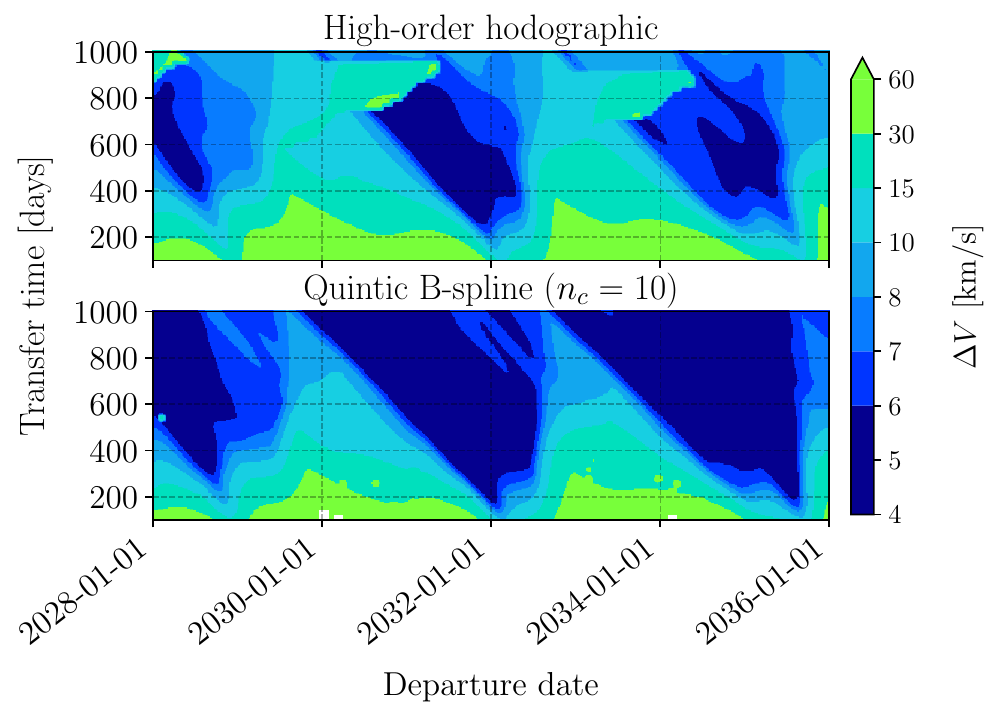}
	\caption{Comparison of Earth--1989 ML low-thrust rendezvous
		$\Delta V$ porkchop plots.}
	\label{fig:porkchop1989ML}
\end{figure}

Figure~\ref{fig:porkchop1989ML} shows the corresponding porkchop plots. Their repeating structure is governed by the Earth--1989 ML synodic period of approximately 3.30 years. The quintic B-spline configuration with 10 control points produces lower $\Delta V$ values over most of the grid, particularly for intermediate and long transfer times. The hodographic porkchop is not expected to reproduce Figure~10 in \cite{Gondelach2015} exactly because that figure employs a different higher-order basis from the best Pareto-front configuration summarized in Table~\ref{tab:gondelach_basis}. Specifically, their Figure~10 uses
$\{1,\tau,\tau^2,\tau\sin(\pi\tau),\tau\cos(\pi\tau)\}$ for both $v_r$ and $v_\theta$, and
$\{\cos\omega_N,\tau^3\cos\omega_N,\tau^3\sin\omega_N\}$ for $v_z$.

\begin{figure}[htbp!]
	\centering
	\includegraphics[width=0.9\linewidth]
	{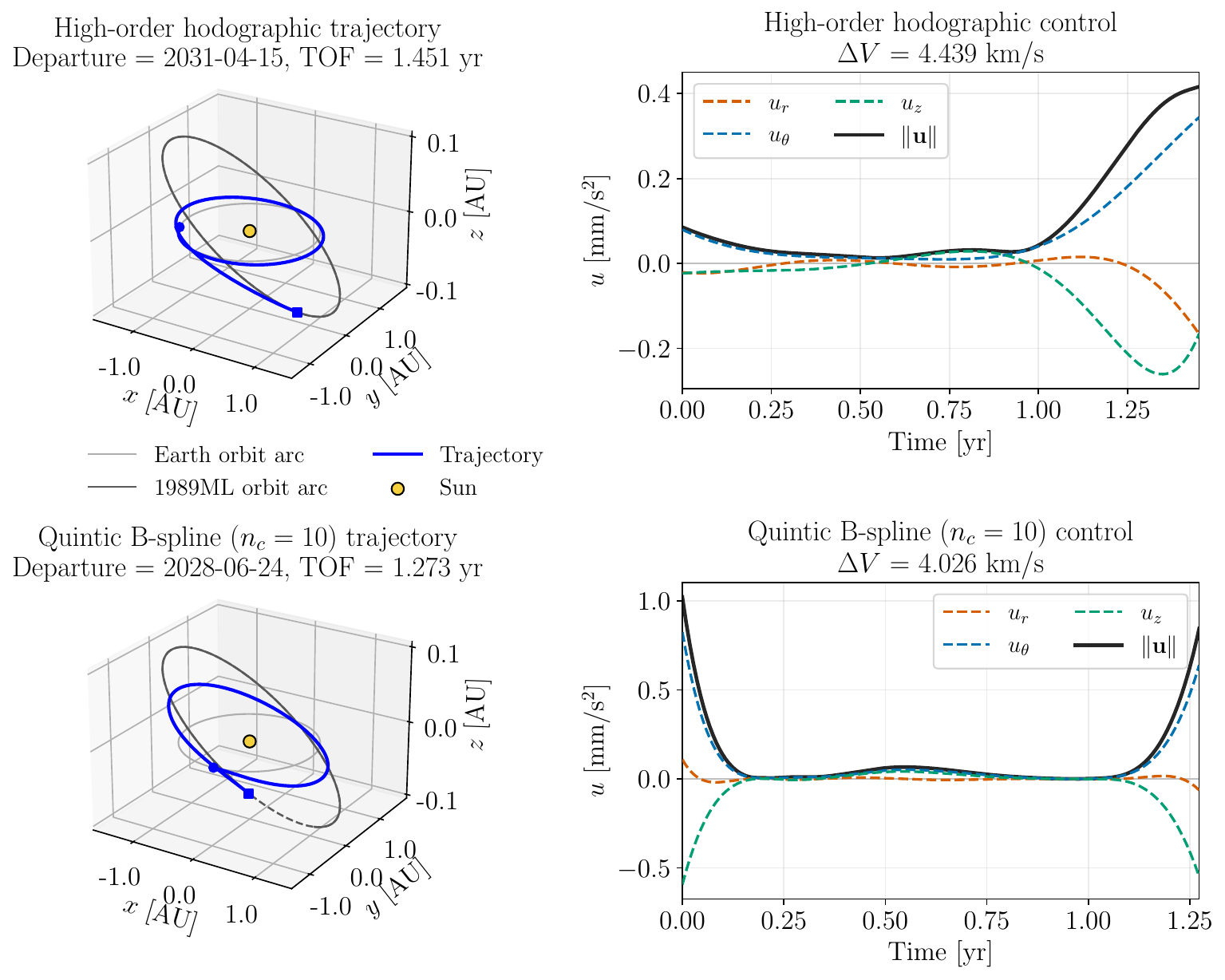}
	\caption{Lowest-$\Delta V$ Earth--1989 ML trajectories and control
		accelerations obtained using high-order hodographic shaping
		and a quintic B-spline with $n_c=10$.}
	\label{fig:1989MLbest_comparison}
\end{figure}

For the refined analysis, the departure window is restricted to the interval from 1 January 2028 to 15 April 2031, which covers approximately one Earth--1989 ML synodic period. Both departure dates and transfer times are sampled at 5-day intervals. Figure~\ref{fig:1989MLbest_comparison} compares the lowest-$\Delta V$ solutions found within the refined grid using high-order hodographic shaping and the quintic B-spline configuration with 10 control points. The corresponding costs are 4.439 and 4.026~km/s, respectively. The B-spline solution therefore reduces $\Delta V$ by approximately 9.3\% relative to the hodographic solution. The hodographic transfer departs on 15 April 2031 and lasts 530 days, whereas the B-spline transfer departs on 24 June 2028 and lasts 465 days. Both solutions correspond to $N=1$. The hodographic control magnitude remains relatively low during the first portion of the transfer and increases substantially near arrival. By contrast, the B-spline solution concentrates most of the control effort near departure and arrival, with a long intermediate interval of nearly vanishing thrust. Its endpoint thrust is primarily divided between the tangential and out-of-plane components.

\begin{figure}[htbp!]
	\centering
	\includegraphics[width=0.6\linewidth]
	{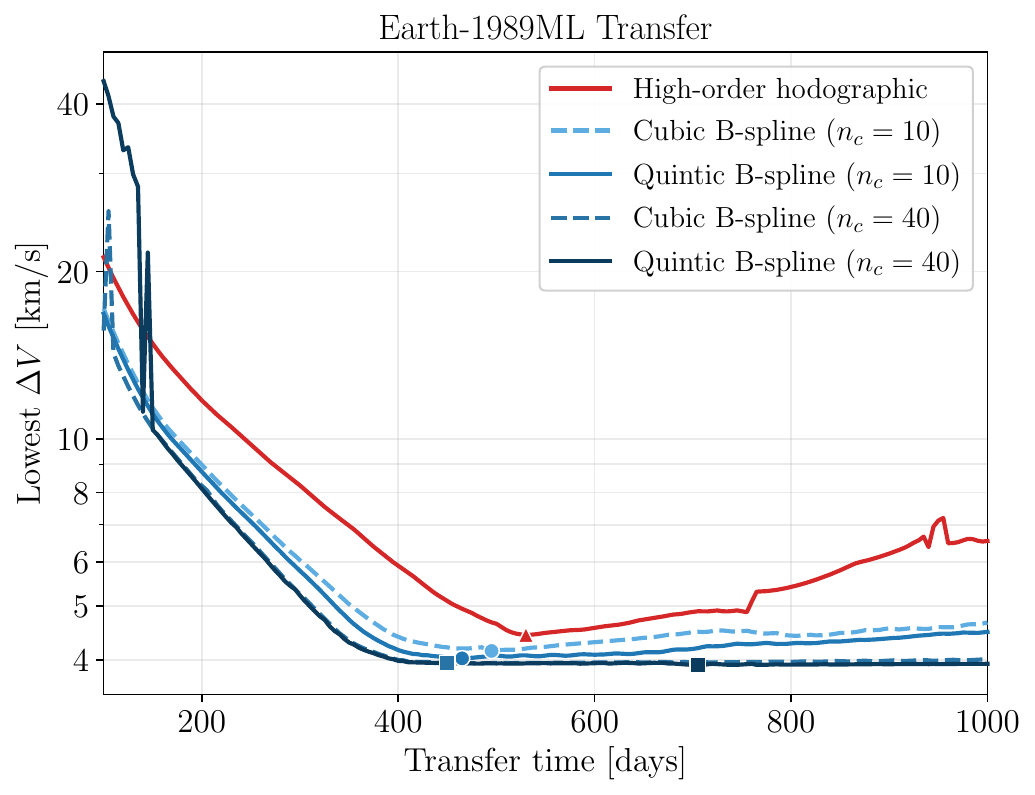}
	\caption{Minimum $\Delta V$ over the departure-date grid for each
		Earth--1989 ML transfer time from 1 January 2028 to 15 April 2031.}
	\label{fig:tofvsfuel1989ML}
\end{figure}

\begin{figure}[htbp!]
	\centering
	\includegraphics[width=0.6\linewidth]
	{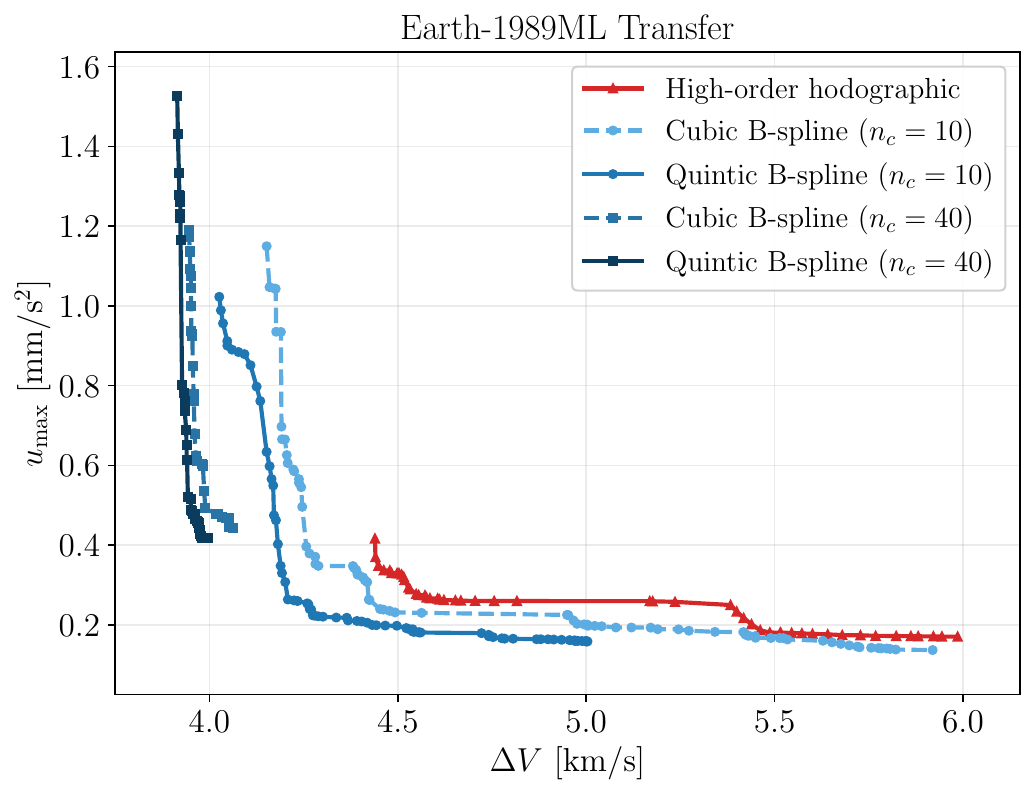}
	\caption{Pareto front of maximum control acceleration versus
		$\Delta V$ for Earth--1989 ML transfers from 1 January 2028
		to 15 April 2031.}
	\label{fig:pareto1989ML}
\end{figure}

The aggregate results are presented in
Figures~\ref{fig:tofvsfuel1989ML} and~\ref{fig:pareto1989ML}. Transfer
times below approximately 200 days require considerably higher
$\Delta V$ for all methods. Beyond approximately 400 days, the
B-spline costs approach 4~km/s and vary only slightly with transfer
time. This behavior is particularly evident for the configurations
with 40 control points, which maintain costs slightly below 4~km/s
over most of the remaining interval. The hodographic solution reaches
its minimum near 530 days and subsequently increases, reaching
approximately 6.6~km/s at 1{,}000 days.

The Pareto fronts show that all B-spline configurations extend toward
lower $\Delta V$ values than the hodographic method. The configurations
with 40 control points attain the lowest costs, although their
lowest-$\Delta V$ solutions require comparatively high maximum control
accelerations. By contrast, the configurations with 10 control points
provide lower values of $u_{\max}$ over a broader portion of the cost
range. Consequently, no single B-spline configuration dominates the
complete trade-off between $\Delta V$ and maximum control acceleration.

\FloatBarrier
\subsubsection{Comet Tempel 1}

Comet Tempel~1 has a comparatively high orbital eccentricity,
$e\approx0.46$, and an inclination of approximately $10.5^\circ$, as
shown in Table~\ref{tab:target_elements}. Departure dates from
1 January 2028 to 28 December 2043 and transfer times from 400 to
1{,}500 days are considered, with both grids sampled at 20-day
intervals. The revolution branches are
$N\in\{0,1,2,3,4,5\}$.

\begin{figure}[htbp!]
	\centering
	\includegraphics[width=0.7\linewidth]
	{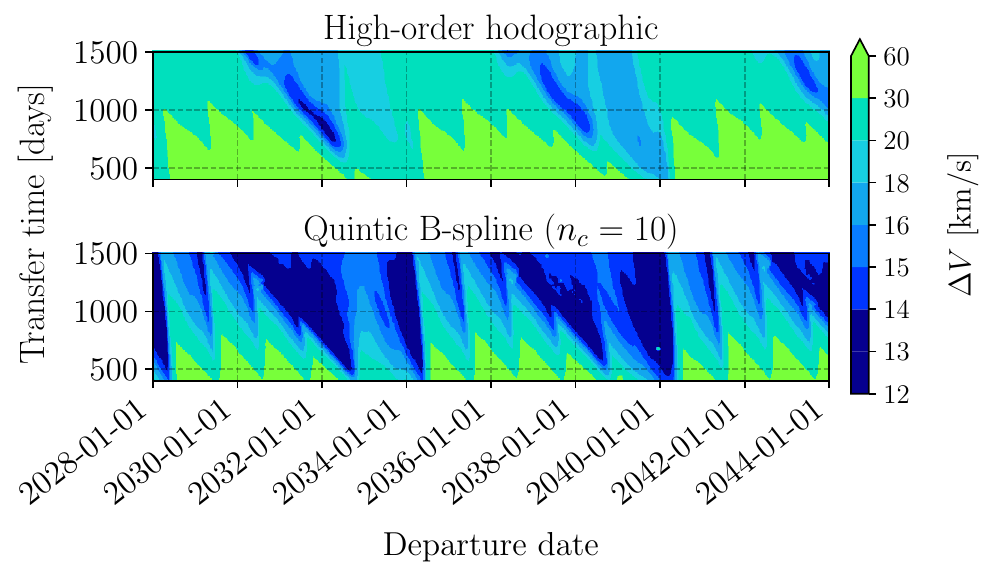}
	\caption{Comparison of Earth--Tempel 1 low-thrust rendezvous
		$\Delta V$ porkchop plots.}
	\label{fig:porkchopTempel1}
\end{figure}

Figure~\ref{fig:porkchopTempel1} shows the corresponding porkchop plots. Similar transfer opportunities occur at intervals of approximately six years, consistent with the orbital period of Tempel~1. The B-spline method produces lower $\Delta V$ values over a considerably larger portion of the departure-date and transfer-time domain than the high-order hodographic method.

\begin{figure}[htbp!]
	\centering
	\includegraphics[width=0.9\linewidth]
	{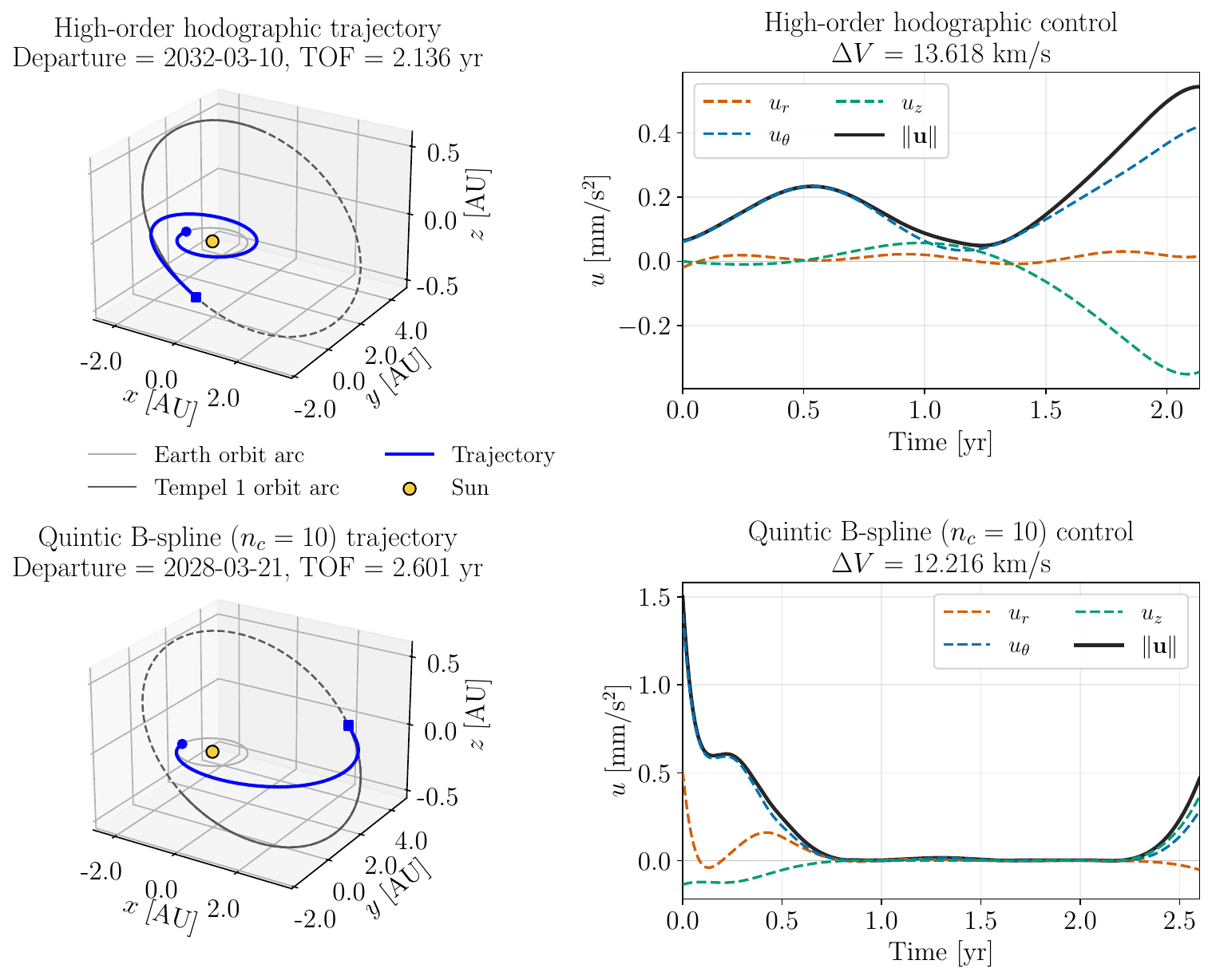}
	\caption{Lowest-$\Delta V$ Earth--Tempel 1 trajectories and control
		accelerations obtained using high-order hodographic shaping
		and a quintic B-spline with $n_c=10$.}
	\label{fig:Tempel1best_comparison}
\end{figure}

For the refined analysis, the departure window is restricted to the interval from 1 January 2028 to 30 December 2033, which covers approximately one orbital period of Tempel~1. Both departure dates and transfer times are sampled at 10-day intervals. Figure~\ref{fig:Tempel1best_comparison} compares the lowest-$\Delta V$ solutions found using high-order hodographic shaping and the quintic B-spline configuration with 10 control points. The corresponding costs are 13.618 and 12.216~km/s, respectively. The B-spline solution therefore reduces $\Delta V$ by approximately 10.3\% relative to the hodographic solution, although it requires a longer transfer. The hodographic transfer departs on 10 March 2032 and lasts 780 days, whereas the B-spline transfer departs on 21 March 2028 and lasts 950 days. The hodographic and B-spline solutions correspond to $N=1$ and $N=0$, respectively. The B-spline solution applies most of its control effort near departure, followed by an extended intermediate coast and a smaller increase near arrival. The hodographic control remains nonzero throughout the transfer and increases substantially during its final portion. In both solutions, the tangential component provides most of the acceleration magnitude, while the axial component becomes significant near arrival.

\begin{figure}[htbp!]
	\centering
	\includegraphics[width=0.6\linewidth]
	{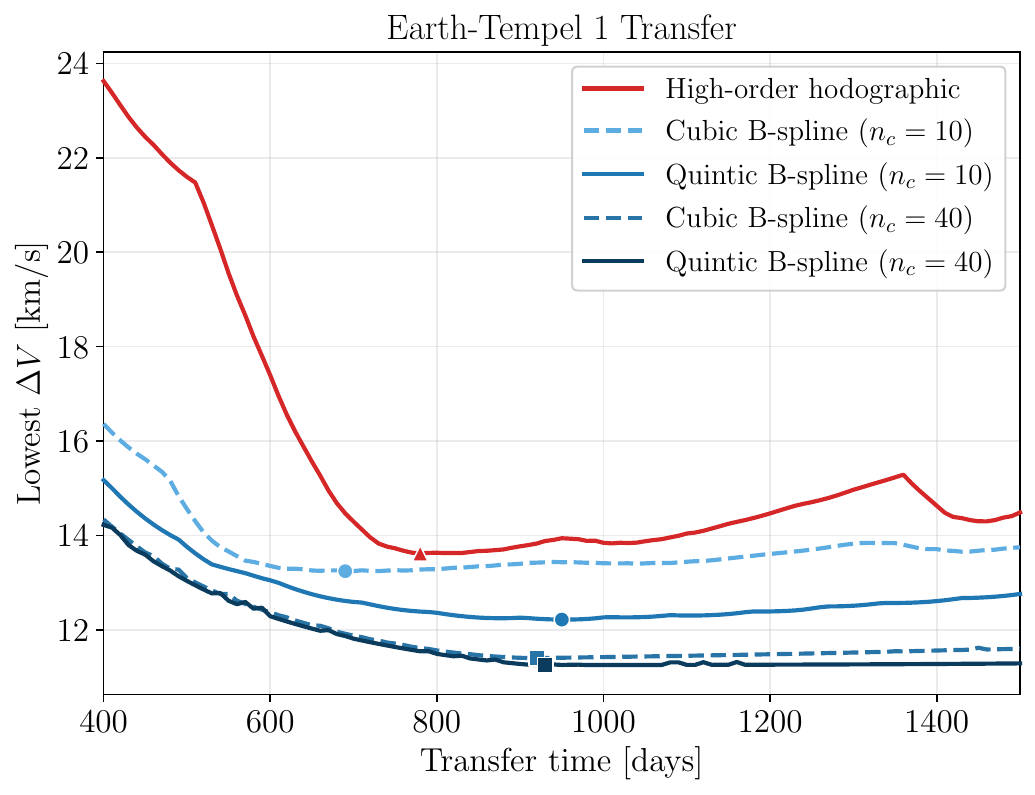}
	\caption{Minimum $\Delta V$ over the departure-date grid for each
		Earth--Tempel 1 transfer time from 1 January 2028 to
		30 December 2033.}
	\label{fig:tofvsfuelTempel1}
\end{figure}

\begin{figure}[htbp!]
	\centering
	\includegraphics[width=0.6\linewidth]
	{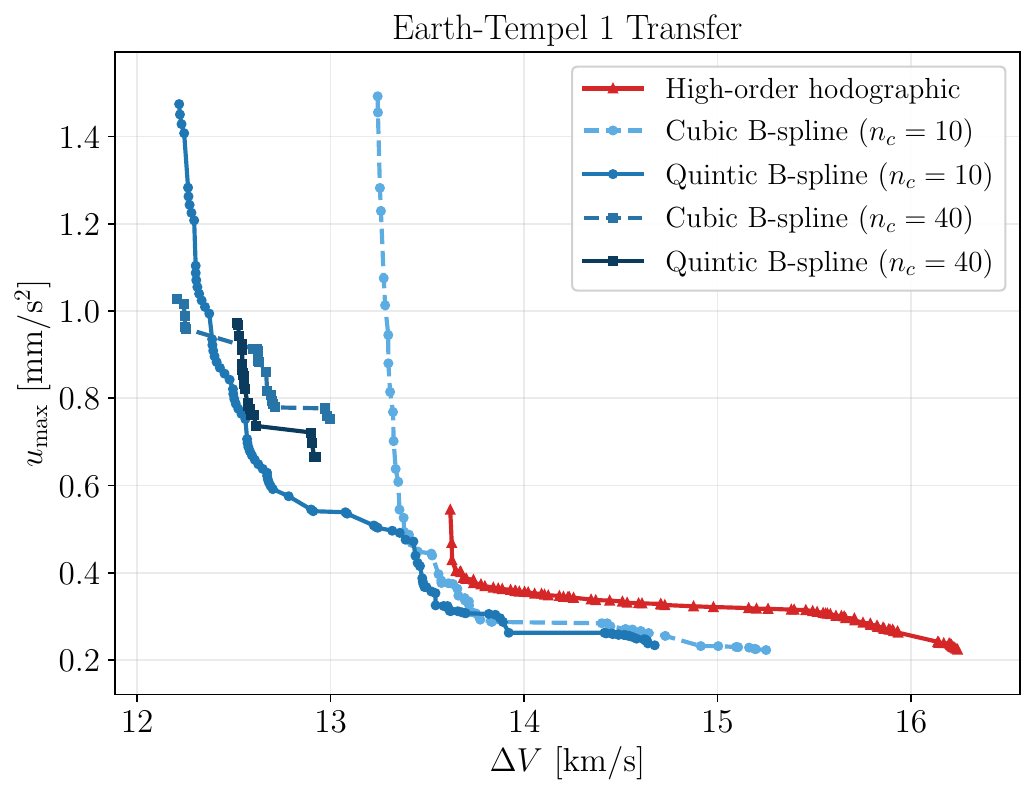}
	\caption{Pareto front of maximum control acceleration versus
		$\Delta V$ for Earth--Tempel 1 transfers from 1 January 2028
		to 30 December 2033.}
	\label{fig:paretoTempel1}
\end{figure}

The aggregate results are presented in Figures~\ref{fig:tofvsfuelTempel1} and~\ref{fig:paretoTempel1}. Transfer times below approximately two years require considerably higher $\Delta V$ for most methods. Beyond this range, the B-spline costs become nearly constant, particularly for the configurations with 40 control points. The hodographic cost reaches its minimum near 780 days and then gradually increases.

The Pareto fronts exhibit the expected trade-off between $\Delta V$ and maximum control acceleration. The configurations with 40 control points attain the lowest $\Delta V$ values, although generally at comparatively high maximum accelerations. The cubic and quintic configurations with 10 control points cover a wider range of $u_{\max}$ and therefore provide additional trade-offs between control authority and propellant consumption.

\FloatBarrier
\subsection{Solution quality and computational efficiency}

Table~\ref{tab:fine_grid_delta_v_statistics} summarizes the global
$\Delta V$ statistics obtained over the fine-grid campaigns for the four targets. The mean
and median characterize the finite solutions retained across each tested
grid, whereas the minimum represents the best transfer opportunity found
for each method and target.

\begin{table*}[htb!]
	\caption{Global $\Delta V$ statistics for the fine-grid transfer campaigns.}
	\label{tab:fine_grid_delta_v_statistics}
	\centering
	\begin{tabular}{llrrr}
		\toprule
		Target & Method & \multicolumn{3}{c}{$\Delta V$ [km/s]} \\
		\cmidrule(lr){3-5}
		& & Mean & Median & Minimum \\
		\midrule
		Mars & High-order hodographic          & 8.184  & 7.758  & 5.762 \\
		Mars & Cubic B-spline ($n_c=10$)       & 7.116  & 6.693  & 5.727 \\
		Mars & Quintic B-spline ($n_c=10$)     & 6.498  & 6.068  & 5.681 \\
		Mars & Cubic B-spline ($n_c=40$)       & 6.056  & 5.692  & 5.630 \\
		Mars & Quintic B-spline ($n_c=40$)     & 6.071  & 5.655  & 5.591 \\
		\addlinespace
		Mercury & High-order hodographic       & 73.489 & 49.774 & 19.823 \\
		Mercury & Cubic B-spline ($n_c=10$)    & 64.529 & 38.156 & 19.400 \\
		Mercury & Quintic B-spline ($n_c=10$)  & 44.616 & 27.199 & 18.864 \\
		Mercury & Cubic B-spline ($n_c=40$)    & 21.256 & 19.593 & 18.093 \\
		Mercury & Quintic B-spline ($n_c=40$)  & 21.172 & 19.198 & 17.566 \\
		\addlinespace
		Tempel 1 & High-order hodographic      & 26.873 & 23.332 & 13.618 \\
		Tempel 1 & Cubic B-spline ($n_c=10$)   & 20.355 & 17.130 & 13.241 \\
		Tempel 1 & Quintic B-spline ($n_c=10$) & 19.218 & 15.958 & 12.216 \\
		Tempel 1 & Cubic B-spline ($n_c=40$)   & 17.490 & 13.366 & 11.402 \\
		Tempel 1 & Quintic B-spline ($n_c=40$) & 18.623 & 13.409 & 11.251 \\
		\addlinespace
		1989 ML & High-order hodographic       & 18.911 & 12.670 & 4.439 \\
		1989 ML & Cubic B-spline ($n_c=10$)    & 14.470 & 8.855  & 4.152 \\
		1989 ML & Quintic B-spline ($n_c=10$)  & 13.934 & 7.867  & 4.026 \\
		1989 ML & Cubic B-spline ($n_c=40$)    & 13.299 & 6.693  & 3.943 \\
		1989 ML & Quintic B-spline ($n_c=40$)  & 15.264 & 6.771  & 3.915 \\
		\bottomrule
	\end{tabular}
\end{table*}

On the tested fine grids and among the finite solutions retained, all
B-spline configurations reduce the mean, median, and minimum $\Delta V$
relative to high-order hodographic shaping. The quintic
B-spline with 10 control points provides a particularly favorable
compromise between solution quality and model complexity. Relative to
the hodographic method, it reduces the median $\Delta V$ by 21.8\% for
Mars, 45.4\% for Mercury, 31.6\% for Tempel~1, and 37.9\% for 1989 ML.
The corresponding reductions in the minimum $\Delta V$ are 1.4\%,
4.8\%, 10.3\%, and 9.3\%, respectively. The median
is more representative of typical grid-wide performance,
whereas the minimum characterizes the best transfer opportunity found.

Increasing the number of control points to 40 provides additional
flexibility and generally improves the best solution found. In
particular, the quintic B-spline with 40 control points achieves the
lowest minimum $\Delta V$ for every target. Relative to the
hodographic method, the reductions are 3.0\% for Mars, 11.4\% for
Mercury, 17.4\% for Tempel~1, and 11.8\% for 1989 ML. The benefit of
B-spline shaping is therefore more pronounced for Mercury, Tempel~1,
and 1989 ML than for Mars.

The computational performance of the methods is summarized in
Table~\ref{tab:fine_grid_computational_times}. Each grid attempt
corresponds to one optimization problem associated with a unique
combination of departure date, transfer time, and number of spacecraft
revolutions $N$. Thus, the number of attempts is the product of the
numbers of departure dates, transfer times, and admissible revolution
branches. For each target, all methods are evaluated over the same set
of attempts.

\begin{table*}[htb!]
	\caption{Computational performance for the fine-grid transfer campaigns.}
	\label{tab:fine_grid_computational_times}
	\centering
	\begin{tabular}{llrrr}
		\toprule
		Target & Method & Grid attempts & Mean [s] & Serial time [h] \\
		\midrule
		Mars & High-order hodographic          & 140{,}430 & 0.158 & 6.17 \\
		Mars & Cubic B-spline ($n_c=10$)       & 140{,}430 & 0.113 & 4.41 \\
		Mars & Quintic B-spline ($n_c=10$)     & 140{,}430 & 0.109 & 4.26 \\
		Mars & Cubic B-spline ($n_c=40$)       & 140{,}430 & 0.846 & 33.02 \\
		Mars & Quintic B-spline ($n_c=40$)     & 140{,}430 & 1.247 & 48.66 \\
		\addlinespace
		Mercury & High-order hodographic       & 141{,}102 & 0.188 & 7.37 \\
		Mercury & Cubic B-spline ($n_c=10$)    & 141{,}102 & 0.115 & 4.50 \\
		Mercury & Quintic B-spline ($n_c=10$)  & 141{,}102 & 0.125 & 4.89 \\
		Mercury & Cubic B-spline ($n_c=40$)    & 141{,}102 & 0.903 & 35.38 \\
		Mercury & Quintic B-spline ($n_c=40$)  & 141{,}102 & 1.665 & 65.25 \\
		\addlinespace
		Tempel 1 & High-order hodographic      & 146{,}520 & 0.167 & 6.81 \\
		Tempel 1 & Cubic B-spline ($n_c=10$)   & 146{,}520 & 0.108 & 4.41 \\
		Tempel 1 & Quintic B-spline ($n_c=10$) & 146{,}520 & 0.113 & 4.60 \\
		Tempel 1 & Cubic B-spline ($n_c=40$)   & 146{,}520 & 0.797 & 32.45 \\
		Tempel 1 & Quintic B-spline ($n_c=40$) & 146{,}520 & 1.211 & 49.27 \\
		\addlinespace
		1989 ML & High-order hodographic       & 130{,}863 & 0.179 & 6.50 \\
		1989 ML & Cubic B-spline ($n_c=10$)    & 130{,}863 & 0.113 & 4.12 \\
		1989 ML & Quintic B-spline ($n_c=10$)  & 130{,}863 & 0.120 & 4.35 \\
		1989 ML & Cubic B-spline ($n_c=40$)    & 130{,}863 & 0.841 & 30.55 \\
		1989 ML & Quintic B-spline ($n_c=40$)  & 130{,}863 & 1.385 & 50.36 \\
		\bottomrule
	\end{tabular}
\end{table*}

Although the grid attempts for the B-spline campaigns are executed in parallel, all methods
are compared using equivalent serial times to ensure a fair comparison
independent of the number of processors assigned to each campaign. All
execution times were measured on a Mac Studio equipped with an Apple
M4 Max processor. The serial time is obtained by summing the execution
times of all grid attempts and therefore differs from the actual
elapsed time of the parallel campaign. The configurations with 10
control points are consistently faster than the high-order hodographic
method. In particular, the quintic B-spline with 10 control points
requires between 0.109 and 0.125~s per attempt, approximately 30--33\%
less than the hodographic method. Its complete campaigns require
between 4.26 and 4.89~h of equivalent serial computation.

The $\Delta V$ improvement obtained with 40 control points comes at a
substantial computational cost. The cubic configurations with 40
control points are approximately seven to eight times slower than their
10-control-point counterparts, whereas the quintic configurations are
approximately 11--13 times slower. Their complete campaign times range
from approximately 31 to 65~h. In addition, only the 40-control-point
configurations require the corresponding 10-control-point solutions as
warm starts; the 10-control-point configurations do not depend on a
lower-resolution B-spline solution. Thus, the practical computational
cost of a 40-control-point campaign also includes the preceding
10-control-point campaign used to generate its initial guesses.
Moreover, the quintic 40-control-point configuration is consistently
slower than the corresponding cubic configuration while providing only
modest additional reductions in the minimum $\Delta V$.

Overall, the quintic B-spline with 10 control points provides the most
balanced configuration when both computational efficiency and solution
quality are considered. It consistently improves upon the hodographic
solutions while requiring less equivalent serial computation. The
configurations with 40 control points are preferable when minimizing
$\Delta V$ within the tested shape spaces is the primary objective and the additional
computational cost, including the generation of the required
10-control-point warm starts, is acceptable.

\FloatBarrier
\subsection{Influence of B-spline hyperparameters on control shape}

\begin{figure}[htbp!]
	\centering
	\includegraphics[width=0.9\linewidth]
	{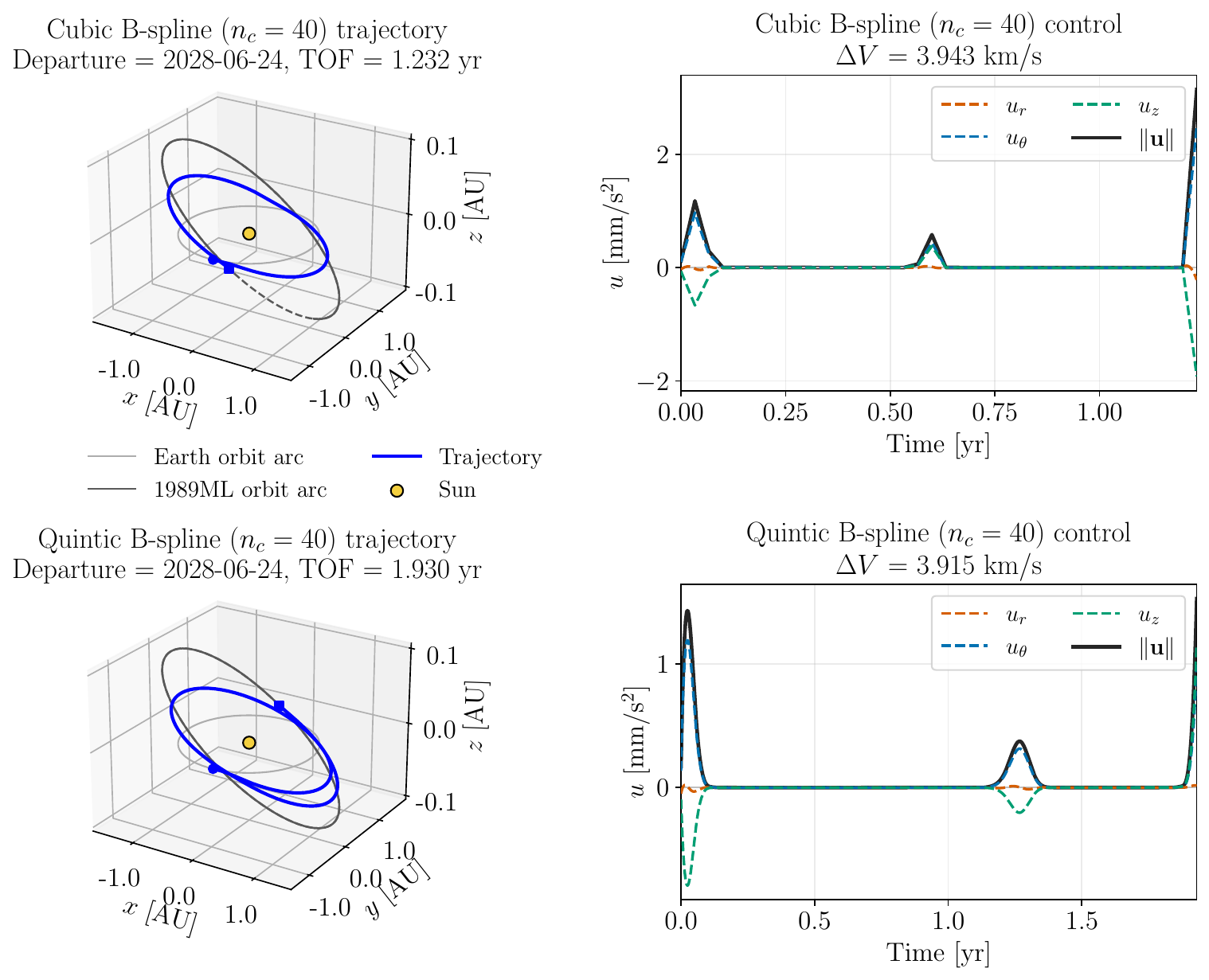}
	\caption{Influence of the B-spline degree on the lowest-$\Delta V$
		Earth--1989 ML trajectories and control accelerations for $n_c=40$.}
	\label{fig:1989ML_control_hyperparameters}
\end{figure}

The B-spline degree and number of control points determine not only the
solution quality but also the regularity and temporal distribution of
the control acceleration. Figure~\ref{fig:1989ML_control_hyperparameters}
shows the lowest-$\Delta V$ Earth--1989 ML solutions obtained using
cubic and quintic B-splines with 40 control points.

For a B-spline trajectory of degree $p$ with simple interior knots, the
trajectory is $C^{p-1}$-continuous. Because the control acceleration
depends on the second derivative of the trajectory, the cubic
B-spline produces a continuous control profile whose slope may be
discontinuous at the knots. These derivative discontinuities generate
the sharp corners and approximately triangular thrust arcs observed in
Figure~\ref{fig:1989ML_control_hyperparameters}.

The quintic B-spline provides greater regularity because its second
derivative is $C^2$-continuous at the interior knots. The resulting
control profile contains smoother transitions between thrust and coast
arcs. With 40 control points, the optimizer concentrates the control
effort into short intervals near departure, at an intermediate point,
and near arrival, while coasting over most
of the transfer. The resulting control history consists of short, concentrated thrust arcs separated by long coast intervals, giving it an impulsive-coast-like structure. This interpretation remains qualitative because the acceleration is continuous, with smooth transitions between finite-duration thrust and coast phases, and no explicit upper bound is imposed on its magnitude.

The effect of the number of control points can also be observed by
comparing the quintic B-spline solution with $n_c=40$ in
Figure~\ref{fig:1989ML_control_hyperparameters} and the $n_c=10$
solution in Figure~\ref{fig:1989MLbest_comparison}. The
10-control-point solution distributes the control over broader portions
of the trajectory, whereas the additional temporal resolution provided
by 40 control points allows the optimizer to produce shorter thrust
arcs and longer coast intervals. Thus, increasing the number of control
points tends to produce a more localized control profile, although it
does not change the theoretical continuity class for a fixed spline
degree. The same qualitative effects of the B-spline hyperparameters on control shape are observed for the other targets.

\section{Conclusions}
\label{sec:conclusions}

This work has presented a B-spline shaping method for fixed-time,
minimum-$\Delta V$ low-thrust interplanetary rendezvous. The method
exploits the differential flatness of the low-thrust dynamics in
cylindrical coordinates to express the required control acceleration uniquely
as a function of the shaped trajectory and its derivatives. A
clamped B-spline parameterization satisfies the position and velocity
boundary conditions analytically, leaving only the interior control
points as optimization variables. The original continuous optimal
control problem is thereby reduced to a finite-dimensional nonlinear program in which the dynamics are satisfied through the flatness mapping. Numerical
quadrature is required only to evaluate the $\Delta V$ objective.

The method was assessed through rendezvous campaigns from Earth to Mars,
Mercury, the near-Earth asteroid 1989~ML, and comet Tempel~1. The B-spline
results were compared with high-order hodographic shaping based on \cite{Gondelach2015}. The same B-spline
architecture was applicable to all four targets without handcrafting
target-specific basis functions. Across the tested fine-grid campaigns
and among the finite solutions retained, every B-spline configuration
reduced the mean, median, and minimum $\Delta V$ relative to the
hodographic benchmark within the respective finite-dimensional shape
spaces. The quintic B-spline
with 10 control points reduced the median $\Delta V$ by 21.8--45.4\%
and the minimum by 1.4--10.3\%, depending on the target. The
improvement was comparatively modest for Mars and more pronounced for
Mercury, asteroid 1989~ML, and comet Tempel~1. Among the analyzed B-spline architectures, the quintic B-spline with 10 control points provided the most favorable compromise between solution quality and computational efficiency. Its
mean execution time ranged from 0.109 to 0.125~s per
departure-date, transfer-time, and revolution-number attempt, which was
approximately 30--33\% lower than that of the hodographic method. The
additional flexibility provided by 40 control points improved the
$\Delta V$ costs but increased the computational time substantially. The 10-control-point quintic formulation is therefore a
suitable default for broad preliminary-design campaigns, whereas the
40-control-point formulations are better suited to the refinement of
selected transfer opportunities.

The conclusions are limited to the finite-dimensional shape spaces and to
the unconstrained-acceleration problem considered in this manuscript. The mass evolution
was neglected, the maximum control acceleration was evaluated only
a posteriori, and the numerical solutions do not constitute proof of
global optimality. However, the shaped trajectories could be
used as rapid initial guesses for higher-fidelity direct or indirect optimal-control
methods. Future work may incorporate explicit thrust models, mass depletion, and
strict acceleration limits into the B-spline shaping formulation.

\subsection*{Acknowledgements}
\label{acknowledgements}

The author gratefully acknowledges financial support from the Spanish Ministry of Science, Innovation and Universities under grant PID2023-147623OB-I00.

\subsection*{Declaration of competing interest}

The author has no competing interests to declare that are relevant to
the content of this article.

\subsection*{Code availability}

The source code used to implement the B-spline shaping method and to generate the numerical results is available at \url{https://github.com/julsanmer/FASTTRANSFER}. 

\section*{References}

 \bibliographystyle{astrobib}
 \bibliography{refs}

\subsection*{Author biography}

\begin{biography}[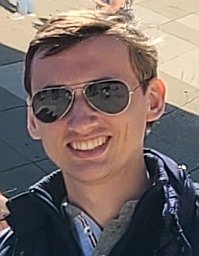]{Julio C. Sanchez}
	is an associate professor in the Department of Aerospace Engineering
	and Fluid Mechanics at Universidad de Sevilla, Spain. Previously, he
	was a Marie Sklodowska-Curie postdoctoral fellow in the Autonomous
	Vehicle Systems Laboratory at the University of Colorado Boulder. He
	received his B.S., M.S., and Ph.D. degrees in aerospace engineering
	from Universidad de Sevilla. His research interests include
		astrodynamics; guidance, navigation, and control; and machine learning.
\end{biography}


\end{document}